\documentstyle[onecolumn]{mn}
\newcommand{\etal}{{et al.}~}

\newcommand{\de}{\delta}
\newcommand{\p}{\partial}
\newcommand{\f}{\frac}

\newcommand{\Lam}{\Lambda}

\newcommand{\eps}{\epsilon}
\newcommand{\Om}{\Omega}

\newcommand{\s}{\sigma}
\newcommand{\al}{\alpha}
\newcommand{\fde}{\tilde{\delta}}
\newcommand{\fpsi}{\tilde{\psi}}

\newcommand{\fW}{\widetilde{W}}

\newcommand{\bfx}{{\bf x}}
\newcommand{\bfr}{{\bf r}}

\newcommand{\bfk}{{\bf k}}

\newcommand{\bfq}{{\bf q}}
\newcommand{\bfg}{{\bf g}}
\newcommand{\bfp}{{\bf p}}
\newcommand{\bft}{{\bf t}}
\newcommand{\bfu}{{\bf u}}
\newcommand{\bfL}{{\bf L}}
\newcommand{\bfS}{{\bf S}}

\newcommand{\vphi}{{\varphi}}

\newcommand{\fvphi}{\tilde{\varphi}}

\newcommand{\calB}{{\cal B}}
\newcommand{\calC}{{\cal C}}
\newcommand{\calD}{{\cal D}}

\newcommand{\calF}{{\cal F}}
\newcommand{\calG}{{\cal G}}

\newcommand{\calJ}{{\cal J}}

\newcommand{\calK}{{\cal K}}

\newcommand{\calT}{{\cal T}}

\newcommand{\bc}{\begin{center}}
\newcommand{\be}{\begin{equation}}
\newcommand{\ee}{\end{equation}}
\newcommand{\ec}{\end{center}}
\newcommand{\lan}{\langle}
\newcommand{\ran}{\rangle}
\newcommand{\sig}{\sigma_{_{\!{\!G}}}}

\def \ltsima{$\; \buildrel < \over \sim \;$}
\def \simlt{\lower.5ex\hbox{\ltsima}}            
\def \gtsima{$\; \buildrel > \over \sim \;$}
\def \simgt{\lower.5ex\hbox{\gtsima}}            
\title[Evolution of the tidal spin from non-Gaussian initial conditions]
{Non-linear evolution of the tidal angular momentum
of protostructures II: non-Gaussian initial conditions} 

\author[P. Catelan and T. Theuns]
{Paolo Catelan$^{12}$ and Tom Theuns $^1$\\
$^1$ Department of Physics, Astrophysics, University of Oxford, Keble Road, 
Oxford OX1 3RH, UK \\
$^2$ Theoretical Astrophysics Center, Juliane Maries Vej 30, DK-2100
Copenhagen, Denmark
} 

\begin{document}

\maketitle

\begin{abstract}
The formalism that describes the non-linear growth of the angular
momentum $\bfL$ of protostructures from tidal torques in a Friedmann
Universe, as developed in a previous paper, is extended to include
non-Gaussian initial conditions. We restrict our analysis here to a
particular class of non-Gaussian primordial distributions, namely
multiplicative models. In such models, strongly correlated phases are
produced by obtaining the gravitational potential via a nonlinear local
transformation of an underlying Gaussian random field. The dynamical
evolution of the system is followed by describing the trajectories of
fluid particles using second-order Lagrangian perturbation theory. In
the Einstein-de Sitter universe, the lowest-order perturbative
correction to the variance of the linear angular momentum of collapsing
structures grows as $t^{8/3}$ for generic non-Gaussian statistics,
which contrasts with the $t^{10/3}$ growth rate characteristic of
Gaussian statistics. This is a consequence of the fact that the
lowest-order perturbative spin contribution in the non-Gaussian case
arises from the third moment of the gravitational potential, which is
identically zero for a Gaussian field. Evaluating these corrections at
the maximum expansion time of the collapsing structure, we find that
these non-Gaussian and non-linear terms can be as high as the linear
estimate, without the degree of non-Gaussianity as quantified by
skewness and kurtosis of the density field being unacceptably large.
The results suggest that higher-order terms in the perturbative
expansion may contribute significantly to galactic spin which contrasts
with the straightforward Gaussian case.
\end{abstract}


\section{Introduction}
It is often assumed that protostructures formed as primordial density
fluctuations are amplified by gravity. The initial density field is
generally assumed to be a Gaussian random field and the amplitude of
fluctuations on all scales is then characterised by the density power
spectrum. An appealing consequence of this gravitational instability
model is that patches of matter are spun-up, as density inhomogeneities
exert tidal torques on collapsing objects (Hoyle 1949), thereby
providing an explanation for the spin of galaxies. The amount of
angular momentum tidally induced on a given patch of matter depends
mainly on its shape and scale $R$ (mass $M$), and to a lesser extent
also on its environment. Simple arguments based on linear gravitational
instability theory lead to values for the angular momentum of the Milky
Way which compare favourable to observational estimates (Peebles
1969). The {\it statistical} distribution of the spin of collapsing
objects has been obtained by studying tidal torques on (high) peaks of
the linear Gaussian density field (Hoffman 1988; Heavens \& Peacock
1988), in the formalism of Peacock and Heavens (1985) and Bardeen \etal
(1986; BBKS). More recently, we adopted the Lagrangian approach
suggested by White (1984) and the BBKS peaks formalism to derive the
probability distribution of the rms spin $L$ (Catelan \& Theuns 1996a):
we showed that the distribution of this variable is practically
indistinguishable from that of the modulus $|\bfL|$ as obtained by
Heavens \& Peacock (1988), indicating that the rms spin is well suited
to characterise the statistics of rotational properties. In addition,
we used the same formalism to derive the distribution of the specific
angular momentum $L/M$ versus $M$ on galactic scales (Catelan \& Theuns
1996a), which allows a direct comparison with the observational
data reported in Fall (1983) as well as with the numerical simulations
of Navarro, Frenk \& White (1995).

The above investigations have been restricted to the linear regime,
during which the galaxy spin grows proportional to the cosmic time $t$
(Doroshkevich 1970; White 1984). We have carried out a theoretical
analysis of how the tidal angular momentum evolves during the mildly
non-linear regime (Catelan \& Theuns 1996b; Paper~I). The formalism
employs perturbative solutions of the Lagrangian dynamical equations
describing the motion of the fluid, and assumes Gaussian initial
conditions. The Lagrangian approach shows to be very well suited to
treat the non-linear evolution of the galactic spin, because it is
powerful in describing the growth of the mass-density fluctuations on
the one hand (Zel'dovich 1970a, b; Buchert 1992; Bouchet \etal 1992;
Catelan 1995); and on the other hand, the usual difficulty of inverting
the mapping from initial Lagrangian coordinates $\bfq$ to Eulerian
coordinates $\bfx$ is completely by-passed. The latter is because the
angular momentum $\bfL$ is invariant with respect to an Eulerian or
Lagrangian description. In spite of the fact that the leading spin
corrections grow faster ($\propto t^{5/3}$) than the linear term, the
main result of the perturbative investigation as reported in Paper~I is
that the predictions of linear theory are rather accurate in
quantifying the evolution of the angular momentum of protostructures
before collapse sets in: non-linear corrections increase the rms spin
by a factor $\sim 1.3$ for the CDM power spectrum on galactic scales.

In this paper we extend the perturbative investigation of Paper~I by
considering non-Gaussian initial conditions. Specifically, we want to
address the question whether allowing the primordial matter field to
deviate from the Gaussian distribution modifies the subsequent
evolution of the angular momentum during the mildly non-linear regime.

A first problem we have to face is the wide variety of possible
non-Gaussian probability distributions.  Primordial non-Gaussian
fluctuations are predicted in many different cosmological
scenarios. Topological defects remaining after an early
phase-transition (Kibble 1976) such as cosmic strings (Zel'dovich 1980;
Vilenkin 1981; Turok 1984; Scherrer, Melott \& Bertschinger 1989),
monopoles (Bennet \& Rhie 1990) or global textures (Davis 1987; Turok
1989; Turok \& Spergel 1990) are examples of models whose statistics
are not Gaussian. Modifications of the inflationary scenario leading to
phase correlations on cosmologically relevant scales have been
discussed (Allen, Grinstein \& Wise 1987; Kofman \& Pogosyan 1988;
Salopek \& Bond 1991; Salopek 1992): these are non-Gaussian as
well. Following suggestions of Kofman \etal (1989), Moscardini \etal
(1991) and Weinberg \& Cole (1992) considered non-Gaussian models
obtained by performing non-linear transformations of an underlying
Gaussian random field. In particular, this last class of non-Gaussian
models is interesting in that one is able, by tuning free model
parameters, to reproduce some of the particular properties of, e.g.,
the texture-seeded Cold Dark Matter model (Gooding, Spergel \& Turok
1991; Park, Spergel \& Turok 1991) or the cosmic explosions scenario
(Ostriker \& Cowie 1981).

Non-Gaussian perturbations constitute a more general statistical model
than Gaussian ones and can be adopted to compute cosmological
observables (see Wise 1988), such as spatial galaxy correlation
functions (Matarrese, Lucchin \& Bonometto 1986; Scherrer \&
Bertschinger 1991), peculiar velocity correlation functions (Scherrer
1992; Catelan \& Scherrer 1995; Moessner 1995), expected size and
frequency of high density regions (Catelan, Lucchin \& Matarrese 1988;
Matsubara 1995), hotspots and coldspots in the cosmic microwave
background radiation (Coles \& Barrow 1987; Kung 1993), and higher
order temperature correlation functions (Gangui \etal 1994; Moessner,
Perivolaropoulos \& Brandenberger 1994).

We limit our analysis in this paper to $multiplicative$ non-Gaussian
processes, which have been analysed recently with $N$-body simulations
in the context of a Cold Dark Matter cosmology (Messina et al. 1990;
Moscardini et al. 1991; Matarrese et al. 1991; Weinberg \& Cole 1992).
In such models, the gravitational potential is obtained via a local
non-linear transformation of an underlying Gaussian random field. These
models are interesting because the procedure provides strongly
correlated phases, while at the same time preserving the standard form
of the initial power spectrum. Consequently, Gaussian and non-Gaussian
models with {\it identical power spectra} can be compared to draw
conclusions on the effects of departing from the more usual assumption
of a normally distributed primordial random field.  Inflation-generated
non-Gaussian fluctuations are expected to be of the multiplicative type
(Matarrese, Ortolan \& Lucchin 1989; Kofman \etal 1990; Barrow \& Coles
1990). An important example of a multiplicative process is the
lognormal statistic, which has recently been proposed as a simple
phenomenological model for describing the (present day) non-linear
density field (Coles \& Jones 1991; see also Bernardeau \& Kofman
1995).

In Section~2 we review how Lagrangian perturbation theory can be used
to obtain corrections to the linear angular momentum due to
non-Gaussian initial conditions. In Section~3 we introduce the
non-Gaussian statistics that we will investigate. In the next Section,
we combine the obtained results to compute the spin corrections for the
different non-Gaussian statistics. Finally, Section~5 summarises our
findings. Technical Appendices contain details on different aspects of
the computations.

\section{Non-linear evolution of the tidal angular momentum}
Let us assume that the behaviour of matter on scales smaller than the
horizon is similar as that of a Newtonian pressureless and irrotational
self--gravitating fluid embedded in an expanding universe with
arbitrary density parameter $\Om$. For the sake of simplicity, we
consider the case of zero cosmological constant and leave to the
interested reader the corresponding generalisation.  Additionally,
luminous objects like galaxies and clusters are assumed to grow through
gravitational amplification of primordial positive density fluctuations
$\de$ of the collisionless fluid.

In what follows, comoving coordinates are denoted by $\bfx$ and
physical distances by $\bfr = a(t)\bfx$, where $a(t)$ is the scale
factor and $t$ the standard cosmic time. The dynamical fluid equations
in a non-flat Friedmann universe simplify considerably if, instead of
$t$, one adopts the temporal coordinate $\tau$ defined by $ d\tau
\equiv a^{-2}\,dt\, $ (Shandarin 1980). In fact, in terms of $\tau$,
the peculiar velocity and the peculiar acceleration are given by $
d\bfx/d\tau\equiv \dot{\bfx}\equiv a(\tau)\,\bfu(\bfx, \tau)\,,$ and
$d^2\bfx/d\tau^2\equiv \ddot{\bfx}\equiv \bfg(\bfx,\tau)\,.$ As
discussed in Shandarin (1980), this dimensionless time $\tau$ is
negative and the initial cosmological singularity at $t=0$ corresponds
to $\tau=-\infty$. Furthermore, the infinity of the cosmic time,
$t=+\infty$, in the open models corresponds to $\tau=-1$, whereas
$t=+\infty$ in the critical Einstein-de Sitter universe corresponds to
$\tau=0$. Finally, the contraction phase in the closed models starts at
$\tau=0$. In terms of the density parameter $\Om$, one has $
\tau=-\sqrt{-k\,}\,(1-\Om)^{-1/2} $, where $k$ is the curvature
constant ($k=-1$ for open universes and $k=1$ for closed
universes). The case $\Om=1\,(k=0)$ is a singular point for the latter
transformation and in this case we take $\tau\equiv-(3t)^{-1/3}$, which
corresponds to using $a(t)\equiv (3t)^{2/3}$ or $t_0=t/a^{3/2}\equiv
1/3$, so defining the unit of time. The scale factor $a(\tau)$ may then
be written for all Friedmann models as $a(\tau)=(\tau^2+k)^{-1}$.

The linear evolution of angular momentum of proto-objects is most easily
analysed using the Zel'dovich (1970a, b) formulation (see White 1984;
Catelan \& Theuns 1996a) and the mildly non-linear spin growth is most
easily analysed using Lagrangian perturbation theory. We recall that
the Zel'dovich approximation coincides with the linear Lagrangian
description.

We will essentially adopt the formulation of the Lagrangian
gravitational theory for a collisionless Newtonian fluid as presented
in Catelan (1995; see also references therein) but note that in the
present paper, as we did in Paper~I as well, the variable $\tau$ has
opposite sign and the growth factor of the linear density perturbation
is normalised differently. An alternative formulation of the Lagrangian
theory may be found in Buchert (1992).

\subsection{Basic tool: Lagrangian theory}
In the Lagrangian formulation, the departure at time $\tau$ of mass
elements from their initial position $\bfq$ is described in terms of
the displacement vector field $\bfS$,
\begin{equation}
\bfx(\bfq, \tau) \equiv \bfq+\bfS(\bfq, \tau)\;.
\label{1}
\end{equation}
The trajectory $\bfS(\bfq, \tau)$ satisfies the Lagrangian \lq
irrotationality\rq~ condition and the Poisson equation given by
(Catelan 1995)
\begin{equation}
\eps_{\al\beta\gamma}\,
\left[ (1+\nabla\cdot\bfS)\,\de_{\beta\sigma} - S_{\beta\sigma}
+ S^C_{\beta\sigma} \right] \dot{S}_{\gamma\sigma} = 0\;,
\label{2}
\end{equation}
\begin{equation}
\left[ (1+\nabla\cdot\bfS)\,\de_{\al\beta} - S_{\al\beta}
+ S^C_{\al\beta} \right] \ddot{S}_{\beta\al} 
=  \al(\tau)[J(\bfq,\tau)-1]\;,
\label{3}
\end{equation}
respectively, where $\eps_{\al\beta\gamma}$ is the totally
antisymmetric Levi-Civita tensor of rank three ($\eps_{123}\equiv 1$),
the symbol $\de_{\al\beta}$ indicates the Kronecker tensor, and
summation over repeated Greek indices (where $\al = 1, 2, 3$) is
understood. In these equations, $\al(\tau)\equiv 6a(\tau)$ and
$J\equiv1/(1+\de)$ is the determinant of the Jacobian of the mapping
$\bfx\rightarrow\bfq$. This determinant $J$ is non-zero until the
first occurrence of shell-crossing (see, e.g., Shandarin \& Zel'dovich
1989). In addition, $S_{\al\beta}\equiv \p S_{\al}/\p q_{\beta}$,
$\nabla \equiv \nabla_{\bfq}$, and $S^C_{\al\beta}$ denotes the
cofactor of $S_{\al\beta}$. In general, $S_{\al\beta}$ is not a
symmetric tensor: $S_{\al\beta}=S_{\beta\al}$ if, and only if, the
Lagrangian motion is longitudinal in which case $\bfS$ is the gradient
of a potential.

The equations~(\ref{2}) and (\ref{3}) are the complete set of
dynamical equations for the vector field $\bfS$, which describes the
trajectory of massive particles of a collisionless fluid embedded in
an arbitrary Friedmann universe. We briefly summarise their
perturbative solutions (up to second-order in the linear displacement)
in the next subsection.

\subsection{Lagrangian perturbative solutions}
The master equations~(\ref{2}) and (\ref{3}) are of third-order and
non-local in the displacement $\bfS$ (see the discussion in Kofman \&
Pogosyan 1995) and it is undoubtedly very difficult to solve them
rigorously. Perturbative solutions can be found by expanding the
trajectory $\bfS$ in a series of which the leading term corresponds to
the Zel'dovich approximation. Specifically, $\bfS=\Sigma_n\bfS_n$,
where $\bfS_n=O(\bfS_1^n)$ is the $n$-th order approximation. Here, we will
need $\bfS_1$ and $\bfS_2$ only. For brevity we use the symbol
$\Theta(\tau)\equiv {\rm ln}[(\tau-1)/(\tau+1)]^{1/2}$ for the open
universe case $(k=-1)$ and $\Lam(\tau) \equiv {\rm
arctang}(1/\tau)=-i\Theta(i\tau)$ for the closed universe case
$(k=+1)$. We neglect decaying modes.

\subsubsection{First-order approximation:}
\noindent The first-order solution to equations~(\ref{2}) and
(\ref{3}) is separable in space and time and corresponds to the
Zel'dovich approximation (Zel'dovich 1970a, b):
\begin{equation}
\bfS_1(\bfq, \tau)=D(\tau)\,\bfS^{(1)}(\bfq)\equiv 
D(\tau)\,\nabla\psi^{(1)}(\bfq)\;;
\label{5}
\end{equation}
the function $D(\tau)$ is the growth factor of linear density
perturbations and is given by:
\begin{equation}
D(\tau)=\left\{
\begin{array}{ll}\f{5}{2}
\left\{1+3\,(\tau^2-1)\left[1+\tau\,\Theta(\tau)\,\right]\right\} 
&\mbox{if $\Om<1$}\\
\tau^{-2} & \mbox{if $\Om=1$}\\
\f{5}{2}\left\{-1+3\,(\tau^2+1)\left[1-\tau\,\Lam(\tau)\,\right]
\right\} &\mbox{if $\Om>1$\,.}
\end{array}
\right.
\label{6}
\end{equation}
The solution for the closed models can be obtained from that for the
open models by substituting in the latter $\tau$ by $i\tau$ and
reversing the sign to make the growing mode positive. Note that, in
contrast to Bouchet \etal (1992) and Catelan (1995), we normalised
$D(\tau)$ according to the suggestion of Shandarin (1980): the
coefficient $5/2$ is such that $D(\tau)\rightarrow \tau^{-2}$ in the
limit $\tau\rightarrow -\infty$, which coincides with the Einstein-de
Sitter case. $D(\tau)$ for the different universes is plotted in Fig.~1
of Shandarin (1980) and in Fig.~A1 of Paper~I. The function
$\psi^{(1)}(\bfq)$ is the (initial) gravitational potential. For later
use we define its Fourier transform, $ \fpsi^{(1)}(\bfp)= \int
d\bfq\,\psi^{(1)}(\bfq)\,{\rm e}^{-i\bfp\cdot\bfq}\;, $ where $\bfp$ is
the comoving Lagrangian wave vector. The Fourier transform of the
linear density field, $\fde^{(1)}(\bfp, \tau) =D(\tau)\fde_1(\bfp)$, is
related to $\fpsi^{(1)}(\bfp)$ via the Poisson equation,
$\fpsi^{(1)}(\bfp)=p^{-2}\fde_1(\bfp)\;.  $

\subsubsection{Second-order approximation}
\noindent The second-order solution is also separable and describes a
longitudinal motion in Lagrangian space:
\begin{equation}
\bfS_2(\bfq, \tau)=E(\tau)\,\bfS^{(2)}(\bfq)\equiv 
E(\tau)\,\nabla\psi^{(2)}(\bfq)\;.
\label{7}
\end{equation}
The second order growth rate $E(\tau)$ is
\begin{equation}
E(\tau)=\left\{
\begin{array}{ll}
- \f{25}{8} -
\f{225}{8}\,(\tau^2-1)\left\{1+\tau\,\Theta(\tau) +\f{1}{2}\left[\tau
+(\tau^2-1)\,\Theta(\tau)\,\right]^2 \right\} & \mbox{if $\Om<1$}\\
-\f{3}{7}\tau^{-4} & \mbox{if $\Om=1$}\\
- \f{25}{8} +
\f{225}{8}\,(\tau^2+1)\left\{1-\tau\,\Lam(\tau)
-\f{1}{2}\left[\tau -(\tau^2+1)\,\Lam(\tau)\,\right]^2\right\} &
\mbox{if $\Om>1$\,.}
\end{array}
\right.
\label{8}
\end{equation}
The second-order non-flat solution was derived previously by Bouchet
\etal (1992). The extra factor $25/4$ of the present formulation is
due to the different normalisation of the first-order solution $D$. An
excellent approximation of the second-order growing mode is $E\approx
-\f{3}{7}D^2$ (see e.g. Fig.~5 of paper~I). In the limit
$\tau\rightarrow -\infty$ one has $E=-\f{3}{7}\tau^{-4}$ which
corresponds to the flat case. The Fourier transform of the
second-order potential $\psi^{(2)}$ is (Catelan 1995):
\begin{equation}
\fpsi^{(2)}(\bfp) 
=-\f{1}{p^2}\int \f{d\bfp_1 d\bfp_2}{(2\pi)^6} 
\big[(2\pi)^3 \de_D(\bfp_1+\bfp_2 - \bfp)\big]\,
\kappa^{(2)}(\bfp_1,\bfp_2)
\,\fpsi^{(1)}(\bfp_1)\,\fpsi^{(1)}(\bfp_2)\;,
\label{9}
\end{equation}
where we have defined the symmetric kernel
\begin{equation}
\kappa^{(2)}(\bfp_1,\bfp_2)\equiv 
\f{1}{2}\,\big[\,p_1^2\,p_2^2 - (\bfp_1\!\cdot\bfp_2)^2\big] \;,
\label{10}
\end{equation}
which describes the second order non-linear corrections
to the trajectory of the fluid elements.\\

The perturbative expansion for $\bfS$ corresponds to a Taylor series
in the variable $D(\tau)=\tau^{-2}$ in the Einstein-de Sitter
universe but this is no longer rigorously true in a non-flat universe.
However, since the higher-order growth factors can be approximated
exceedingly well by powers of $D$, the expansion in the non-flat case
is still \lq close\rq~ to a Taylor expansion (see the discussion in
Paper~I). We proceed by considering how corrections to the linear
displacement $\bfS_1$ translate into corrections to the linear angular
momentum.

\subsection{Non-linear spin dynamics}
In this section, we briefly summarise the perturbative approach to the
spin evolution. As stressed in Paper~I, the angular momentum of the
matter contained in the volume $V(\tau)$ at a given time $\tau$ may be
written equivalently as an integral over the corresponding {\it
initial} volume $\Gamma$:
\begin{equation}
\bfL(\tau) = \eta_0\,\int_\Gamma d\bfq\;[\bfq+\bfS(\bfq, \tau)]
\times\f{d\bfS(\bfq, \tau)}{d\tau}\;.
\label{lin}
\end{equation}
Here, $\eta_0\equiv a^3\rho_b$, where $\rho_b$ is the mean background 
density. This procedure enables us to apply the Lagrangian description
of Newtonian gravity previously reviewed. The linear regime
(Zel'dovich approximation) has been fully analysed in this way by
Doroshkevich (1970), White (1984) and Catelan \& Theuns (1996a),
whereas its Eulerian counterpart was studied extensively by Heavens
and Peacock (1988). We can extend the linear Lagrangian analysis of the
evolution of the angular momentum $\bfL(\tau)$ to the non-linear
regime by applying perturbation theory to equation~(\ref{lin}). Perturbative
corrections to $\bfS(\bfq, \tau)$ (Bouchet \etal 1992; Buchert 1994;
Catelan 1995 and references therein) then give perturbative
corrections to $\bfL(\tau)$: formally
\begin{equation}
\bfL(\tau)=\sum_{h=0}^\infty\,\bfL^{(h)}(\tau)\equiv
\sum_{h=0}^\infty\,
\sum_{j=0}^h \,\eta_0\!\int_\Gamma d\bfq\, \,\bfS_j(\bfq, \tau)\times
\f{d \bfS_{h-j}(\bfq, \tau)}{d\tau}\;,
\label{11}
\end{equation}
with $\bfS_0\equiv\bfq$, hence $\bfL^{(0)}={\bf 0}$. To calculate the
lowest-order corrections to the ensemble average $\lan \bfL^2\ran$,
we need to compute corrections to $\bfL$ up to second-order. After
reviewing briefly the results of the linear theory, we summarise the
final expression of the correction $\bfL^{(2)}$.

\subsubsection{Linear approximation}
\noindent The linear Lagrangian theory corresponds to the Zel'dovich
approximation and the first-order term in equation~(\ref{11}) is given
by:
\begin{equation}
\bfL^{(1)}(\tau)=
\eta_0\,\dot{D}(\tau)\int_{\Gamma}d\bfq\,\bfq\times\nabla\psi^{(1)}(\bfq)\;.
\label{l1}
\end{equation}
Assuming that $\psi^{(1)}(\bfq)$ can be adequately represented in the
volume $\Gamma$ by the first three terms of the Taylor series about
the origin, $\bfq={\bf 0}$, each component $L^{(1)}_{\al}(t)$ may be
written in compact form as (White 1984; Catelan \& Theuns 1996a):
\begin{equation}
L^{(1)}_{\al}(t)=\dot{D}(\tau)\,\eps_{\al\beta\gamma}\,
\calD^{(1)}_{\beta\s}\,\calJ_{\s\gamma}=
-\dot{D}(\tau)\,\eps_{\al\beta\gamma}\,\calJ_{\s\gamma}
\int\f{d\bfp}{(2\pi)^3}\,p_{\s}\,p_{\gamma}\,\fW(pR)\,\fpsi^{(1)}(\bfp)\;,
\label{13}
\end{equation}
where we introduced the deformation tensor at the origin, $
\calD^{(1)}_{\beta\s}\equiv\calD^{(1)}_{\beta\s}({\bf 0})=
\p_{\beta}\p_{\s}\psi^{(1)}({\bf 0})$, and the inertia tensor of the
mass contained in the volume $\Gamma$, $
\calJ_{\s\gamma}\equiv\eta_0\int_{\Gamma}d\bfq\,q_{\s}\,q_{\gamma}\;.
$ In addition, the field $\psi^{(1)}$ is now assumed to be filtered on
scale $R$ using the smoothing function $W_R$, whose Fourier transform
is $\fW(pR)$. Equation~(\ref{13}) shows that the linear angular
momentum $\bfL^{(1)}$ is in general non--zero because the principal
axes of the inertia tensor $\calJ_{\al\beta}$, which depend only on
the (irregular) shape of the volume $\Gamma$, are not aligned with the
principal axes of the deformation tensor $\calD^{(1)}_{\al\beta}$,
which depend on the location of neighbour matter fluctuations. The
temporal growth of the tidal angular momentum induced by this
misalignment is completely contained in the function $\dot{D}(\tau)$,
which behaves as $\dot D(\tau)=-2\tau^{-3}\sim t$ in the Einstein--de
Sitter universe, as first noted by Doroshkevich (1970). Finally, if
$\Gamma$ is a spherical Lagrangian volume, then
$L^{(1)}_{\al}\sim\eps_{\al\beta\gamma}\,\calD^{(1)}_{\beta\gamma}=0$.
Consequently, the matter contained initially in a spherical volume
does not gain any tidal spin during the linear regime (see also the
discussion in White 1984).

\subsubsection{Second-order approximation}
\noindent The second-order term in equation~(\ref{11}) involves the
second-order displacement $\bfS^{(2)}$:
\begin{equation}
\bfL^{(2)}(\tau)=\eta_0 \int_\Gamma d\bfq\,\bfq\times\f{d\bfS_2}{d\tau}=
\eta_0\,\dot{E}(\tau)\int_{\Gamma}d\bfq\,\bfq\times\nabla\psi^{(2)}(\bfq)\;.
\label{l2}
\end{equation}
Note that, since $E\propto\tau^{-4}$, one
has $\dot{E}\propto\tau^{-5}$ hence the second-order terms grows
$\propto t^{5/3}$ in the Einstein--de Sitter universe. This growth
rate was first derived by Peebles (1969). If we represent
$\psi^{(2)}(\bfq)$ in $\Gamma$ by the first three terms of a Taylor
series, as we did before for $\psi^{(1)}$, we obtain for the
$\al$-component (see Paper~I):
\begin{eqnarray}
L^{(2)}_{\al}(\tau)&=&\dot{E}(\tau)\,\eps_{\al\beta\gamma}\,
\calD^{(2)}_{\beta\s}\,\calJ_{\s\gamma}
\nonumber \\
&=&\dot{E}(\tau)\,\eps_{\al\beta\gamma}\,\calJ_{\s\gamma}
\int\f{d\bfp_1\,d\bfp_2}{(2\pi)^6}\,
\f{(\bfp_1+\bfp_2)_\s\,(\bfp_1+\bfp_2)_\gamma}{|\bfp_1+\bfp_2|^2}\,
\fW(|\bfp_1+\bfp_2|R)\,\kappa^{(2)}(\bfp_1, \bfp_2)\,
\fpsi^{(1)}(\bfp_1)\,\fpsi^{(1)}(\bfp_2)\;,
\label{15}
\end{eqnarray}
where $ \calD^{(2)}_{\beta\s}\equiv\calD^{(2)}_{\beta\s}({\bf 0})=
\p_{\beta}\p_{\s}\psi^{(2)}({\bf 0})\; $ is the second-order
deformation tensor. In addition, the second-order potential
$\psi^{(2)}$ in equation~(\ref{15}) is now explicitely smoothed on
scale $R$ by the filter $\fW(pR)$. The filtering of the field
$\psi^{(2)}$ on an appropriate scale reflects the restriction to the
mildly non-linear evolution of the proto-object, since the strongly
non-linear couplings between different modes are filtered out. Note
that the non-linear dynamical evolution modifies only the deformation
tensor and not the inertia tensor. Furthermore, if $\Gamma$ is a
sphere, then again $\bfL^{(2)}={\bf 0}$. This is in contrast to
Eulerian perturbation theory since the angular momentum of an Eulerian
sphere does grow in second-order perturbation theory (Peebles 1969;
White 1984).\\

For a single collapsing region enclosed in a volume $\Gamma$ it is
enough to evaluate equation~(\ref{11}) at the time of maximum expansion
$\tau_M$ to compute its final angular momentum. After $\tau_M$ the
angular momentum essentially stops growing since the collapsed object
is much less sensitive to external tidal couplings (Peebles
1969). However, highly non-linear interactions typically occuring after
the maximum expansion time may lead to a significant redistribution of
angular momentum in the final object in a complicated way (see
discussion in Catelan \& Theuns 1996a and references therein). 

In order to compare the theory against statistical results obtained
from $N$--body simulations or from observations, it is useful to
compute the variance of the angular momentum of the object for an {\it
ensemble} of realisations of the gravitational potential random field
$\psi^{(1)}$. This programme is carried out in the next section.

\section{Non-Gaussian random fields: spin ensemble averages}
We simplify the previous results by considering the expectation value
over the ensemble of realisations of the non-Gaussian random field
$\psi^{(1)}$ of the square of $\bfL$, $\lan \bfL^2\ran$. Note that one
can relatively easily obtain the probability distribution for the {\em
linear} angular momentum (Heavens \& Peacock 1988; Catelan \& Theuns
1996a). However, such a calculation would be far more complicated for
the non-linear contributions, even in the Gaussian case. We will follow
the same procedure adopted in Paper~I, i.e.  we neglect the
correlations between the inertia tensor and the gravitational field
$\psi^{(1)}$.

Taking into account the mildly non-linear corrections one has
\begin{equation}
\lan\bfL^2\ran = 
\lan\bfL^{(1)2}\ran + 2\lan\bfL^{(1)}\!\cdot\bfL^{(2)}\ran
+O(\tau^{-9})\;.
\label{lensem}
\end{equation}
The linear term $\lan\bfL^{(1)2}\ran$ has been discussed extensively by
Catelan \& Theuns (1996a). Since it involves only the second-order
moment of the underlying distribution (i.e. the power spectrum), its
formal expression is valid for all gravitational potential fields, both
Gaussian and non-Gaussian. We report its explicit expression below (see
Eq.~49).  Furthermore, it is easy to understand from
equations~(\ref{13}) and (\ref{15}) that
$\lan\bfL^{(1)}\!\cdot\bfL^{(2)}\ran$ is the lowest-order perturbative
correction since it corresponds to an integral over the bispectrum
$B_\psi^{(1)}$ of the initial gravitational potential field
$\psi^{(1)}$:
\begin{equation}
(2\pi)^3\,\de_D(\bfp_1+\bfp_2+\bfp_3)\, B_\psi^{(1)}(\bfp_1, \bfp_2)
\equiv
\lan\fpsi^{(1)}(\bfp_1)\,\fpsi^{(1)}(\bfp_2)\,\fpsi^{(1)}(\bfp_3)\,
\ran\;.
\label{bispec}
\end{equation}
Here, $\delta_D$ is the Dirac delta function. Specifically, one gets
\begin{equation}
\lan\bfL^{(1)}\!\cdot\bfL^{(2)}\ran=
\dot{D}(\tau)\,\dot{E}(\tau)\,\eps_{\al\beta\gamma}\,
\eps_{\al\beta'\gamma'}\,
\calJ_{\s\gamma}\,\calJ_{\s'\gamma'}\,
\lan\,\calD^{(1)}_{\beta\s}\calD^{(2)}_{\beta'\!\s'}\ran\;,
\label{18}
\end{equation}
where
\begin{eqnarray}
\lan\,\calD^{(1)}_{\beta\s}\calD^{(2)}_{\beta'\!\s'}\ran&=&\!\!
-\int\f{d\bfp_1\,d\bfp_2}{(2\pi)^6}\,
\f{(\bfp_1+\bfp_2)_\beta\,(\bfp_1+\bfp_2)_\s\,
(\bfp_1+\bfp_2)_{\beta'}\,(\bfp_1+\bfp_2)_{\s'}}{|\bfp_1+\bfp_2|^2}\,
\kappa^{(2)}(\bfp_1,
\bfp_2)\,\big[\fW(|\bfp_1+\bfp_2|R)\big]^2\,B_\psi^{(1)}(\bfp_1,
\bfp_2) \nonumber \\ 
&\equiv& \calK^{(2)}_{\beta\s\beta'\!\s'}\oplus
B_\psi^{(1)}\;,
\label{}
\end{eqnarray}
which defines the kernel $\calK^{(2)}_{\beta\s\beta'\!\s'}$ describing
the effects of the 2-order dynamics. To maintain a compact notation,
we have introduced the operation ``$\oplus$'' defined for integrable
functions $\calF$ and $\calG$ as
\begin{equation}
\calF\oplus\calG
\equiv \int\f{d\bfp_1 d\bfp_2}{(2\pi)^6}\calF(\bfp_1, \bfp_2)\,
\calG(\bfp_1, \bfp_2)\;.
\end{equation}
Saturating the tensors' indices in equation (\ref{18}) and using the
isotropy of the universe, it is possible to write equation (\ref{18})
in full generality as (see Appendix~A for details)
\begin{equation}
\lan\bfL^{(1)}\cdot\bfL^{(2)}\ran = \f{2}{15}\,
\dot{D}(\tau)\dot{E}(\tau)\,(\mu_1^2-3\mu_2)\,\Sigma(R)\;,
\label{21}
\end{equation}
where
\begin{equation}
\Sigma(R) \equiv 15\,\lan\calD^{(1)}_{xy}\,\calD^{(2)}_{xy}\ran =
15\,\calK^{(2)}_{xyxy}\oplus B_\psi^{(1)}\;.
\end{equation}
Here, $\mu_1$ and $\mu_2$ are the first and the second invariants of
the inertia tensor $\calJ$ (section 3.2 below). The factorisation in
terms of the invariant $\mu_1^2-3\mu_2$ is typical of spin ensemble
averages, both in linear and mildly non-linear regime, as far as one
investigates the rotational properties of objects with preselected
inertia tensor. We remark that the lowest-order perturbative correction
to $\lan\bfL^{(1)2}\ran$ grows as $\dot{D}(\tau)\,\dot{E}(\tau)\propto
\tau^{-8}\propto t^{8/3}$ which contrasts with the Gaussian case where
the growth rate is $t^{10/3}$ instead (cfr. Paper~I).

Given a particular probability distribution for the linear
gravitational potential $\psi^{(1)}$, one can compute the explicit
value of $\Sigma(R)$, which involves an integral over the
bispectrum. Alternatively, one can use Poisson's equation to obtain a
relation between the potential and density bispectrum and compute the
integral over the density bispectrum $B_\de^{(1)}$ instead. Indeed,
Poisson's equation implies the following relation between the two
spectra
\begin{equation}
B_\de^{(1)}(\bfp_1, \bfp_2)=
p^2_1\,p^2_2\,|\bfp_1+\bfp_2|^2\, B_\psi^{(1)}(\bfp_1, \bfp_2)\;.
\label{}
\end{equation}
The latter expression is useful whenever a non-Gaussian distribution is
assumed directly for the density field, rather than for the underlying
gravitational potential field. For example, an interesting such case is
the lognormal density distribution which was recently proposed by Coles
and Jones (1991) as a reliable empirical fit for the present day
density distribution function.

Before performing the detailed calculations of the non-linear
corrections to the spin, we need to describe the properties of those
non-Gaussian fields which we intend to investigate. Unfortunately, this
next section turns out to be rather long and readers not interested in
the murky details can skip directly to subsection~3.2, where the final
expressions for the spin averages are summarised.

\subsection{Non-Gaussian initial conditions}
We will consider non-Gaussian statistics that are obtained from
non-linear local transformations of an underlying Gaussian random
field. These statistics are assumed either for the potential or
directly for the density field and we will start with the first case.
We restrict ourselves to the same non--Gaussian statistics numerically
explored by Moscardini et al.  (1991; see also Matarrese et al. 1991;
Messina et al. 1992; Moscardini et al.  1993), namely the {\em
lognormal} (hereafter LN) and the {\em Chi--squared} (hereafter
$\chi^2$) model. These statistics are imposed on the primordial
potential, i.e. before the linear transfer function has transformed the
initial scale free power spectrum to its present day shape. As an
example, we will specialise to the Cold Dark Matter (CDM) transfer
function, although our method is not restricted to CDM.

This primordial potential is a gauge--invariant (up to a zero--mode)
variable whose statistics are in principle fixed by primordial
processes, e.g. in the inflationary scenario. This potential is not
required to have zero mean, since only its gradient is physically
meaningful. This allows for a wider variety of underlying statistics
than if the probability distribution were to be imposed directly on
the fluctuating density field, since the latter is required to have
zero mean and in addition needs to be positive, $\rho>0$, in each point.

The models we consider are obtained by the following procedure:
$\psi^{(1)}$ is found by convolving a stationary random field
$\vphi(\bfq)$ (the primordial potential) with a real function
$\calT(\bfq)$ (the transfer function):
\begin{equation}
\psi^{(1)}(\bfq)\equiv \int d\bfq' \, \calT(\bfq' - \bfq)\, \varphi(\bfq')\;.
\label{24}
\end{equation}
The primordial potential $\vphi$ is calculated by applying some
non--linear operation on an underlying Gaussian random process $G$,
e.g. $\vphi\propto G^2$ for $\chi^2$ and $\vphi\propto \exp(G)$ for LN
statistics.  The transfer function $\calT(\bfq)$ is characterised by
its Fourier transform:
\begin{equation}
\calT(\bfp) \equiv \int d\bfq\,\calT(\bfq)\,{\rm e}^{-i\bfp\cdot\bfq}
\equiv T(p)\,F_\vphi(p)^{-1/2}\;,
\label{25}
\end{equation}
where $T(p)$ is the CDM transfer function (e.g. Efstathiou 1990)
\begin{equation}
T(p)=\left[1+\big(ap +(bp)^{3/2}+ (cp)^2\big)^\nu\right]^{-1/\nu}\;,
\end{equation}
with $a=6.4\,(\Om h^2)^{-1}$ Mpc,~$b=3.0\,(\Om h^2)^{-1}$ Mpc$,~
c=1.7\,(\Om h^2)^{-1}$ Mpc, and $\nu=1.13$. The function $F_\vphi(p)$
is a positive correction factor applied to obtain the required CDM
power spectrum for $\psi^{(1)}$ when the power spectrum of $\vphi$ is
assumed to be scale free:
\begin{equation}
P_\psi(p)= \big[\calT(p)\big]^2\,P_\vphi(p) \equiv 
A\,p^{-3}\,\big[T(p)\big]^2\;,
\end{equation}
and $P_\psi(p)\sim p^{-3}$ for $p\rightarrow 0$. The latter relation
also defines the normalisation constant $A$, which has to be
appropriately calculated in each model since it depends on the
underlying spectrum $P_\vphi(p)$. We will show that it is possible to
factorise the dependence as $P_\vphi(p)\approx A_\vphi
F_\vphi(p)\,p^{-3}$, from which we obtain $A \approx
A_\vphi\,F_\vphi(p)^{-1}F_\vphi(p)=A_\vphi$, where the
scale-dependence introduced by the function $F_\vphi$ cancels out.
Consequently, the normalisation of $\psi$ follows directly from the
amplitude of $\vphi$.

The assumption implicit in equation (\ref{24}) is that the statistics
of $\psi^{(1)}$ are of primordial origin, i.e., they refer to the time
when the fluctuations were outside the Hubble radius. In addition, the
overall sign has to be fixed by the primordial physical mechanism
inducing the fluctuation field. The latter statistic is left invariant
in the linear regime, independent of the subsequent processes which
modify the wave content of $\psi^{(1)}$ on all scales smaller than the
Hubble radius.

Note that both $\chi^2$ and LN statistics belong to the same general
class $\varphi({\bf q}) \propto \vert 1 + \alpha^{-1} G({\bf q})
\vert^\alpha$; in fact, for $\alpha=2$ and $\lan G^2 \ran^{1/2}\equiv
\sig \gg 1$, the $\chi^2$ distribution is recovered, while $\alpha\to
\infty$ yields the lognormal one.  For small wave numbers, where
$\calT(p)\approx 1$, the gravitational potential of scale--invariant
distributions obeys the scaling law: 
\begin{equation}
\lan \fpsi^{(1)}(\mu \bfp_1)
\cdots \fpsi^{(1)}(\mu \bfp_N)\ran \,d(\mu \bfp_1) \cdots d(\mu\bfp_N)
\approx \lan \fpsi^{(1)}(\bfp_1) \cdots \fpsi^{(1)}(\bfp_N)\ran
\,d\bfp_1 \cdots d\bfp_N, 
\end{equation} 
for any $N$, up to logarithmic corrections which arise as a
consequence of the necessity to perform these integrations over a
finite $\bfp$ volume. This gives a straightforward generalisation of
the Harrison--Peebles--Zel'dovich scale--invariance to non--Gaussian
statistics.

To compute the non-linear evolution of the tidal angular momentum we
need to compute the primordial bispectrum $B_\psi^{(1)}$, which in
turn is determined by the bispectrum $B_\vphi$, defined as
usual. Since $\fpsi^{(1)}(p)= \calT(p)\,\fvphi(p)$, we find a relation
between the ``transferred'' and the primordial potential bispectra:
\begin{equation}
B_\psi^{(1)}(\bfp_1,\bfp_2) = \calT(p_1)\,\calT(p_2)\,
\calT(|\bfp_1+\bfp_2|)\,B_\vphi(\bfp_1, \bfp_2)\;.
\label{29}
\end{equation}
It should be stressed again that the density field and the
gravitational potential do not, in general, obey the same
statistics. In addition, one cannot self-consistently assume any
non-Gaussian distribution for the density field $\de_1$: for example, a
Chi-squared density distribution has a positive definite 2-point
correlation function, at variance with the assumed large-scale
homogeneity. In contrast, the same statistics for $\vphi$ do not lead
to an inconsistency. Let us now describe the various models in more
detail.

\subsubsection{Chi--squared model}
The $\chi^2$ distribution for $\vphi$ is obtained from the following
transformation of the underlying Gaussian field $G$:
\begin{equation}
\vphi(\bfq) = \vphi_\circ\, G(\bfq)^2\;,
\end{equation}
where $G$ has zero-mean and variance $\sig^2$. The power spectrum of
the field $G$ is chosen as
\begin{equation}
P_G(p) = p^{-3}\,\Pi(p)\;.
\label{31}
\end{equation}
Here we introduced the operator
$\Pi(\bfp)=\Pi(|\bfp|)=\vartheta(p-p_m)\, \vartheta(p_M-p)$, where
$\vartheta(x)$ denotes the Heaviside step function. The result of
projecting with $\Pi(\bfp)$ in equation~(\ref{31}) is that
$P_G(p)=p^{-3}$ in the interval $[p_m,\,p_M]$ and zero outside. The
cutoff constants $p_m$ and $p_M$ are introduced to avoid divergences at
both infra-red and ultra-violet \lq wavelengths\rq~. In terms of these
cutoffs, the field $G$ has variance
$\sig^2=(2\pi^2)^{-1}\ln(p_M/p_m)$. On the other hand, the constant
$\vphi_\circ$ has the same dimensions as $\vphi$ and can be either
negative ($\chi^2_n$ model) or positive ($\chi^2_p$ model). These names
are chosen according to the sign of the corresponding skewness
$\langle\delta^3\rangle/\langle \delta^2\rangle^{3/2}$: the $\chi^2_n$
($\chi^2_p$) model is negatively (positively) skewed. Finally, it is
worth noting that the flicker-noise choice for the $P_G$ spectrum,
giving equipartition in Fourier space, is stable up to negligible
corrections against the non-linear transformation originating $\vphi$,
both in Chi-square and LN non-Gaussian models (see below).

The Chi-squared model is an example of a scale-invariant non-Gaussian
statistic (Otto et al. 1986; Lucchin \& Matarrese 1988). Such models
have been considered by Coles \& Barrow (1987) in connection with the
two-dimensional distribution of CBR fluctuations.  In the inflationary
context, Bardeen (1980) proposed a model where adiabatic perturbations
are described by a squared Gaussian process, our $\chi^2_p$, whose $G$
field has a non-scale-free power spectrum.

It can be shown that the $\varphi$ power spectrum corresponding to the
choice (\ref{31}) is (see Appendix~B for details)
\begin{equation}
P_\vphi(p) \approx {\vphi_\circ^2 \over 2\pi^2} 
\, \left[ \beta(p) + 2 \ln(1+p/p_m) -1 \right]\,p^{-3}\,\Pi(p)\;,
\label{32}
\end{equation}
where $\beta(p) = (1-p/p_m)^2$, for $p_m \leq p \leq 2p_m$, and
$\beta(p) =1 + 2 \ln(-1+p/p_m)$, for $2p_m \leq p \ll p_M$.  The
normalisation constant $A$ is related to $\vphi_\circ$ as $A\approx
A_\vphi=\vphi_\circ^2/2\pi^2\,$.

Relevant to our investigation is the expression for the linear
gravitational bispectrum $B_\vphi$, which turns out to be given by a
convolution over the underlying spectrum $P_G$, namely (see Appendix~C
for details)
\begin{equation}
B_\vphi(\bfp_1,\bfp_2) = \f{\vphi_\circ^3}{\pi^3} \int
d\bfp\, P_G(p)\, P_G(|\bfp_1 +\bfp|)\,P_G(|\bfp_2-\bfp|)\;.
\label{33}
\end{equation}
Note that the statistics of the Chi-squared potential $\vphi$ are
independent of the variance $\sig$ of the underlying Gaussian field
(see Appendix~C). This is not the case for the LN distribution (see
below), for which the departure from Gaussianity can be tuned by
varying $\sig$.

Finally, we define for later use the operator
\begin{equation}
\Pi(\bfp_1,\bfp_2)=\Pi(\bfp_1)\,\Pi(\bfp_2)\,\Pi(\bfp_1+\bfp_2)\;,
\label{eq:theta}
\end{equation}
i.e., the projection $\Pi(\bfp_1,\bfp_2)$ is zero if any of the two
vectors $\bfp_1$ or $\bfp_2$ or their sum has modulus $\le p_m$ or $\ge
p_M$ and is one otherwise.

\subsubsection{Lognormal model}
The LN distribution for $\varphi$ is obtained from the following
transformation of the underlying Gaussian field $G$:
\begin{equation}
\vphi(\bfq) = \vphi_\circ\, {\rm e}^{G(\bfq)}\;,
\label{34}
\end{equation}
where, as before, $G$ has zero-mean and variance $\sig^2$.  Note that
$\lan\vphi\ran=\vphi_\circ\,\exp(\sig^2/2)$ and there is again the
freedom of assuming $\vphi_\circ$ to be negative (LN$_n$ model) or
positive (LN$_p$ model).

The lognormal statistics has been analysed in the cosmological context
by many authors (see e.g., Coles 1989; Coles \& Jones 1990; Messina et
al. 1990; Coles , Melott \& Shandarin 1993; Catelan \etal 1994; Sheth
1995).

In contrast to the $\chi^2$ case, one gets lognormal statistics which
differ by varying amounts from the Gaussian distribution by tuning the
variance of $G$: for small $\sig$, the non--Gaussian character of
$\varphi$ is manifest only in the properties of rare high peaks,
whereas for larger $\sig$, the power spectrum $P_\vphi$ deviates
strongly from flicker--noise.

Next, we analyse the hierarchical structure of the correlation functions
of the field $\vphi$. The 2-point correlation function 
$\lan\vphi(\bfq_1)\,\vphi(\bfq_2)\ran -\lan\vphi\ran^2\equiv\xi_\vphi(q)$
is given by
\begin{equation}
\xi_\vphi(q)=
\lan\vphi\ran^2\,\left[{\rm e}^{\xi_G(q)}-1\right]\;,
\end{equation}
where $\xi_G(q) \equiv \lan G(\bfq_1)\,G(\bfq_2)\ran$ and $q\equiv
|\bfq_1-\bfq_2|$. The 3-point correlation function $\lan
(\vphi_1-\lan\vphi\ran)\, (\vphi_2-\lan\vphi\ran)\,
(\vphi_3-\lan\vphi\ran)\ran\equiv\zeta_\vphi(123)$ is (see Appendix~C
for the derivation)
\begin{equation}
\zeta_\vphi(123) 
=\lan\vphi\ran^{-1}
\big[\xi_\vphi(12)\,\xi_\vphi(13)+\xi_\vphi(12)\,\xi_\vphi(23)+
\xi_\vphi(13)\,\xi_\vphi(23)\big] + 
\lan\vphi\ran^{-3}\big[\xi_\vphi(12)\,\xi_\vphi(13)\,\xi_\vphi(23)\big]\;.
\end{equation}
The power spectrum $P_\vphi(p)$ is given by
\begin{equation}
P_\vphi(p) = \lan\vphi\ran^2\,
\int d\bfq \left[{\rm e}^{\xi_G(q)}-1\right]\,{\rm e}^{i\bfp\cdot\bfq}\;,
\label{37}
\end{equation}
and the formal expression for the bispectrum is
\begin{equation}
\lan\vphi\ran^3 B_\vphi(\bfp_1,\bfp_2) =
\lan\vphi\ran^2 \left\{P_\vphi(p_1)\,P_\vphi(p_2) + 
P_\vphi(|\bfp_1+\bfp_2|)\,\big[P_\vphi(p_1)+P_\vphi(p_2)\big]\right\}
+\int \!\f{d\bfp}{(2\pi)^3}\, 
P_\vphi(p)\,P_\vphi(|\bfp_1+\bfp|)\, P_\vphi(|\bfp_2-\bfp|)\,.
\label{38}
\end{equation}
A remark is now appropriate. Since we are interested in comparing the
results of these non-Gaussian investigation against the results from
Gaussian initial conditions reported in Paper~I, it is important to
preserve the form of the power spectrum. On the other hand, it can be
easily understood from equation (\ref{37}) that the spectrum of the
field $\vphi$ is a complicated function of the wavevector $p$, even
for the simple underlying spectrum $P_G(p)$ adopted in equation
(\ref{31}). Fortunately, the transformed field $\vphi$ may be easily
constrained to have the required power spectrum if we restrict
ourselves to those LN distributions for which $\sig^2\leq 1$. This
restriction leads to some loss of generality, but the formalism
becomes much simpler, and, above all, the power spectrum of the
underlying Gaussian field is preserved.  In this case, since
$\xi_G(q)$ is a decreasing function of $q$, the condition $\sig^2\leq
1$ implies that $|\xi_G(q)|\leq 1$, and a Taylor expansion of the
exponential in the integrand (\ref{37}) leads to $ P_\vphi(p) \approx
\lan\vphi\ran^2 \,P_G(p)\;.  $ In the following we restrict ourselves to
LN distributions with $\sig^2 \simlt 1$. Consequently, the power
spectrum $P_\vphi$ may be written as
\begin{equation} 
P_\vphi(p) \approx \f{2\pi^2
\sig^2\,\vphi_\circ^2\,{\rm e}^{\sig^2}} {\ln(p_M/p_m)}\,p^{-3}
\Pi(p)\;,
\end{equation} 
and the bispectrum as
\begin{eqnarray}
B_\vphi(\bfp_1,\bfp_2)&\approx&
\lan\vphi\ran^3 \left[\f{2\pi^2
\sig^2}{\ln(p_M/p_m)}\right]^2 
\Big[p_1^{-3}p_2^{-3}+|\bfp_1+\bfp_2|^{-3}(p_1^{-3}+p_2^{-3})
\nonumber\\
&+&\f{2\pi^2
\sig^2}{\ln(p_M/p_m)} \!\int \f{d\bfp}{(2\pi)^3}\,
p^{-3}\, |\bfp_1 +\bfp|^{-3}\,|\bfp_2-\bfp|^{-3}
\Pi(\bfp_1,\bfp)\Pi(\bfp_2,-\bfp)\Big]\,\Pi(\bfp_1,\bfp_2)\,.
\label{40}
\end{eqnarray}
Note that the bispectrum $B_\vphi(\bfp_1,\bfp_2)$ is zero outside the
region $p_m\leq p_{1,2}\leq p_M$ and $p_m\leq |\bfp_1+\bfp_2|\leq p_M$.
The potential $\psi$ now has the CDM power spectrum with amplitude $A
\approx 2\pi^2\sig^2\,\vphi_\circ^2\,{\rm e}^{\s^2_G}\,/
\ln(p_M/p_m)$. \\

\subsubsection{``Linearised'' lognormal distribution for the density field}
A completely different approach may be adopted if the non-Gaussian
distribution is imposed directly on the linear {\it density}
fluctuation field $\de_1$, rather than on the underlying
potential. Although one is not allowed to choose any non-Gaussian
density distribution, as argued earlier, there are some advantages in
considering models where non-Gaussianity is coded directly in the
statistics of the matter distribution. An interesting case is the LN
density distribution, proposed by Coles \& Jones (1991) as a reliable
empirical fit to the present day density distribution function.  A
theoretical explanation for this fact followed from the analysis of
Bernardeau (1994), who computed the evolution of the one-point
probability distribution function of the cosmic smoothed density
induced by gravitational evolution. Indeed, his results indicate that,
during the mildly non-linear regime, the density distribution is
fairly well approximated by a LN statistic when the spectral index $n$
is close to $-1$. For larger values of $n$ or for smaller values of
the rms $\s_\de$, the resulting LN distribution is close to
Gaussian. The formation of large-scale structure by gravitational
instability from primordial lognormal density fields has been analysed
by Weinberg and Cole (1992) using $N$-body simulations. A further
reason to consider a LN density field is motivated by the recent
extension of the Hoffman-Ribak algorithm (1991) for the construction
of constrained Gaussian random fields by Sheth (1995) to include also
lognormal statistics.

Let us analyse more closely the complex hierarchical structure of the
LN density distribution. In particular we wish to discuss the
approximate expression for the 3-point LN correlation function we are
allowed to adopt in our investigation. Fortunately, as we will see
below, the lognormal distribution permits the constraining of the
deviations of $p(\de)$ from the Gaussian distribution uniquely in
terms of the linear-regime condition. Let us suppose that the
(linear) density field $\rho\equiv \rho_b[1+\de]$ is obtained by the
transformation
\begin{equation}
\rho\equiv \rho_b\, {\rm e}^{G-\sig^2/2}\;,
\label{44}
\end{equation}
where, as before $G$ is an adimensional zero-mean Gaussian random
field with variance $\sig^2$. The mean value is given by
$\lan\rho\ran=\rho_b$. We stress the fact that equation~(\ref{44}) is
mathematically identical to the transformation (\ref{34}), but, since
it involves the density field, it is physically very different.

The 2-point LN correlation function $\xi_\de(q)$ is given by 
\begin{equation}
\xi_\de(q)\equiv \lan\de_1(\bfq_1)\,\de_1(\bfq_2)\ran = 
{\rm e}^{\xi_G(q)}-1\;,
\label{45}
\end{equation}
where $\xi_G(0)=\sig^2$.  Evaluating $\xi_\de(q)$ at zero lag gives the
variance of the LN random field $\de_1$:
\begin{equation}
\s_\de^2 = {\rm e}^{\sig^2} - 1\;.
\label{46}
\end{equation}
We stress that the equations~(\ref{45}) and (\ref{46}) do not contain
any time dependence and combine two concepts, that of ``non-linear
regime'' and that of ``deviation from Gaussianity''. Since for a LN
distribution the $N$-point ($N>2$) functions can be expressed as a
Kirkwood expansion of the 2-point functions (see Coles \& Jones 1991;
also below), the linear-regime condition can be cast simply in the
form $\s_\de^2\ll 1$ which then also implies $\sig^2\ll 1$, i.e. the
primordial LN density distribution with small fluctuations cannot
deviate strongly from Gaussianity. Alternatively, if the primordial LN
density distribution were to deviate strongly from the Gaussian one
then the initial conditions would contain an {\it intrinsic}
non-linear signal, unacceptable on observational grounds. Finally,
since $\xi_G(q)$ is a decreasing function of $q$, $\sig^2\ll 1$
implies $|\xi_G(q)|\ll 1$ and consequently also $|\xi_\de(q)|\ll 1$,
which is important for the subsequent discussion.

We first show that the density power spectrum and the underlying
Gaussian spectrum $P_G(p)$ are similar in the approximation
$\sigma^2_G\ll 1$. Indeed:
\begin{equation}
P_\de(p)\approx P_G(p)\equiv  A\,p\,\big[T(p)\big]^2\,\Pi(p)
\;,
\end{equation}
where $T(p)$ is the CDM transfer function. Additionally, we derive
some properties of the structure of the LN linear potential bispectrum
$B_\de$, which is the Fourier transform of the 3-point correlation function
$\zeta_\de(\bfq_1, \bfq_2, \bfq_3)\equiv \zeta_\de(123)$, 
\begin{equation}
\zeta_\de(123) = \xi_\de(12)\,\xi_\de(13)+\xi_\de(12)\,\xi_\de(23)
+\xi_\de\,\xi_\de(23)+ \xi_\de(12)\,\xi_\de(13)\,\xi_\de(23)\;.
\label{48}
\end{equation}
The important point to note is that the expressions (\ref{45}) and
(\ref{48}) may be simplified if we restrict ourselves to the case
$\sig^2\ll1$, which corresponds to the linear regime, as explained
earlier. Since in this case also $\xi_\de\ll 1$, it seems reasonable
to approximate the 3-point function $\zeta_\de$ by the relation
\begin{equation}
\zeta_\de(123) \approx
\xi_\de(12)\,\xi_\de(13)+\xi_\de(12)\,\xi_\de(23)
+\xi_\de(13)\,\xi_\de(23)\;.  
\end{equation}
The corresponding bispectrum reduces to 
\begin{eqnarray}
B_\de(\bfp_1,\bfp_2) &\approx& 
P_\de(p_1)\,P_\de(p_2) +
P_\de(|\bfp_1+\bfp_2|)\,\big[P_\de(p_1)+P_\de(p_2)\big]
\nonumber \\
&\approx& 
P_G(p_1)\,P_G(p_2)+P_G(|\bfp_1+\bfp_2|)\,\big[P_G(p_1)+P_G(p_2)\big]\;,
\label{50}
\end{eqnarray}
where the cubic terms have been neglected with respect to the
quadratic ones. This approximation allows a considerable
simplification of the calculations, yet the loss of accuracy of the
whole description is small. We stress that the same strategy cannot be
applied to the LN gravitational potential since the constraint
$\s_\vphi \ll 1$ does not imply a restriction to the linear regime.

\subsection{Ensemble averages}
\subsubsection{Linear approximation: $\lan \bfL^{(1)2}\ran$}
\noindent The ensemble average of the rms linear angular momentum,
$\lan\bfL^{(1)2}\ran$, is computed and discussed extensively in
Catelan \& Theuns (1996a). Here, we briefly summarise the main
results, which hold for any underlying probability distribution,
i.e. for both Gaussian and non-Gaussian statistics. We have:
\begin{equation}
\lan\bfL^{(1)2}\ran=\f{2}{15}\dot{D}(\tau)^2(\mu_1^2-3\mu_2)\,\s(R)^2\;,
\label{51}
\end{equation}
where the quantity $\s(R)^2$ is the mass variance on the scale $R$
which is obtained from the power spectrum through $ \s(R)^2\equiv
(2\pi^2)^{-1}\int_0^{\infty} dp\,p^6\,P_{\psi}(p)\fW(pR)^2\;.$ We will
adopt a Gaussian smoothing function,
$\fW(pR)=\exp(-p^2R^2/2)$. Equation~(\ref{51}) holds for any power
spectrum $\lan\fpsi^{(1)}(\bfp_1)\,\fpsi^{(1)}(\bfp_2)\ran_\psi\equiv
(2\pi)^3\,\de_D(\bfp_1+\bfp_2)\,P_{\psi}(p_1)$ and the value of
$\lan\bfL^{(1)2}\ran$ depends on the normalisation of the spectrum. The
general expression~(\ref{51}) is $independent\,$ of the details of the
boundary surface of the volume $\Gamma$ but depends on $\mu_1$ and
$\mu_2$, which are the first and the second invariant of the inertia
tensor $\calJ_{\al\beta}$. Specifically, denoting the eigenvalues of
the inertia tensor by $\iota_1$, $\iota_2$ and $\iota_3$, one has $
\mu_1\equiv\iota_1+\iota_2+\iota_3\; $ and $
\mu_2\equiv\iota_1\iota_2+\iota_1\iota_3+\iota_2\iota_3\; $. For a
spherical volume, $\iota_1=\iota_2=\iota_3$, hence $\lan
\bfL^{(1)2}\ran \propto \mu_1^2-3\mu_2=0$, as we stressed before.

\subsubsection{Higher-order approximation: 
$\lan\bfL^{(1)}\!\cdot\bfL^{(2)}\ran$}
\noindent The calculation of the term
$\lan\bfL^{(1)}\!\cdot\bfL^{(2)}\ran$ takes advantage of the results of
the second-order approximation, and essentially reduces to specialising
the function $\Sigma(R)$ to the particular statistics chosen, since
$\Sigma(R) = 15\,\calK^{(2)}_{xyxy}\oplus B_\psi^{(1)}$, as explained
in the introduction of Section~3. The final expressions, valid for a
CDM power spectrum $P_{\de}(p)\equiv Ap[T(p)]^2 = p^4\,P_{\psi}(p) =
p^4 \,[\calT(p)]^2\,P_\vphi(p)$, where $\calT(p)\equiv
F_\vphi(p)^{-1/2}\,T(p)$ and $P_\vphi(p)\approx
A_\vphi\,F_\vphi(p)\,p^{-3}\,$, may be written for the different
statistics as (compare with equation~\ref{21}):\\

\noindent $\bullet$ For a Chi-squared gravitational field:
\begin{equation}
\lan\bfL^{(1)}\cdot\bfL^{(2)}\ran = \f{2}{15}\,
\dot{D}(\tau)\dot{E}(\tau)\,(\mu_1^2-3\mu_2)\,\Sigma_{\chi^2}(R)\;,
\end{equation}
where the function $\Sigma_{\chi^2}$ depends on the smoothing scale $R$ and
the normalisation of the spectrum $A$, as
\begin{equation}
\Sigma_{\chi^2}(R) = 15\,\calK^{(2)}_{xyxy}(\bfp_1, \bfp_2)\oplus\calT(p_1)\,
\calT(p_2)\,\calT(|\bfp_1+\bfp_2|)\,B_\vphi(\bfp_1, \bfp_2)\;.
\label{}
\end{equation}
The correction $F_\vphi(p)$ is such that [see equation~(\ref{32}) for the symbols]
\begin{equation}
\calT(p) = T(p)\big[\beta(p)+2\ln(1+p/p_m)-1\big]^{-1/2}\;,
\end{equation}
and the normalisation constant is $A\equiv\vphi_\circ^2/2\pi^2$. The
primordial bispectrum may be deduced from equation~(\ref{33}),
\begin{equation}
B_\vphi(\bfp_1, \bfp_2)= \pm (2A)^{3/2}\int
d\bfp\;p^{-3}\,|\bfp_1+\bfp|^{-3}\,|\bfp_2-\bfp|^{-3}
\Pi(\bfp_1,\bfp)\,\Pi(\bfp_2,-\bfp)\,\Pi(\bfp_1,\bfp_2)
\;.
\end{equation}
Typically, these integrals have to be evaluated numerically.  The upper
(lower) sign is for positively (negatively) skewed distributions. Note
that the average $\lan\bfL^{(1)} \cdot\bfL^{(2)}\ran$ factorises the
same invariant $\mu_1^2-3\mu_2$ of the inertia tensor $\calJ$ as
appeared in the linear term [see equation (\ref{51})]: this is a direct
consequence of the isotropy of the universe which is independent of the
underlying statistical properties of the matter distribution.\\

\noindent $\bullet$ For a LN gravitational potential field, the
calculation may be partially done analytically to give
$\Sigma_{LN}(R)$, which is a sum of three terms:
$\Sigma_{LN}(R)=\Sigma_{LN}^{(a)}(R)+\Sigma_{LN}^{(b)}(R)+
\Sigma_{LN}^{(c)}(R)$ which originate from the Kirkwood expansion of
the bispectrum $B^{(1)}_\psi$ in equation~(\ref{38}):
\begin{equation}
\Sigma_{LN}(R)= 15\, \calK^{(2)}_{xyxy}(\bfp_1,\bfp_2)\oplus
T(p_1)\,T(p_2)\, T(|\bfp_1+\bfp_2|)\,B_\vphi(\bfp_1, \bfp_2)\;;
\end{equation}
where
\begin{eqnarray}
B_\vphi(\bfp_1,\bfp_2)\!&\approx& \!\pm A^{3/2}\!
\sqrt{\f{2\pi^2\,\s^2_G}{\ln(p_M/p_m)}} \,\Big[
p_1^{-3}p_2^{-3}+|\bfp_1+\bfp_2|^{-3}(p_1^{-3}+p_2^{-3})\nonumber\\
&+&
\f{2\pi^2\,\s^2_G}{\ln(p_M/p_m)} \!\int
\!\!\f{d\bfp}{(2\pi)^3}\, p^{-3}\, |\bfp_1
+\bfp|^{-3}\,|\bfp_2-\bfp|^{-3}
\Pi(\bfp_1,\bfp)\Pi(\bfp_2,-\bfp)
\Big]\,\Pi(\bfp_1,\bfp_2)\,.
\end{eqnarray}
In this expression, the product $p_1^{-3}p_2^{-3}$ gives rise to the
$\Sigma_{LN}^{(a)}(R)$ term, $\Sigma_{LN}^{(b)}(R)$ corresponds to
$\sim |\bfp_1+\bfp_2|^{-3}(p_1^{-3}+p_2^{-3})$ and finally
$\Sigma_{LN}^{(c)}(R)$ is the loop contribution. In Appendix~D we will
show how these integrals may be partially done analytically.\\

\noindent $\bullet$ For a ``linearised'' LN density field, we obtain in
a similar way as before the function
$\Sigma_{\de}(R)=\Sigma_{\de}^{(a)}(R)+\Sigma_{\de}^{(b)}(R)$, which is
the sum of two contributions which originate from the two terms in the
linear bispectrum $B_\de^{(1)}$ in equation~(\ref{50}):
\begin{equation}
\Sigma_{\de}(R)= 15\,
\calK^{(2)}_{xyxy}(\bfp_1,\bfp_2)\oplus p_1^{-2}p_2^{-2}|\bfp_1+\bfp_2|^{-2}\,
B^{(1)}_\de(\bfp_1, \bfp_2)\;,
\label{56}
\end{equation}
where
\begin{equation}
B^{(1)}_\de(\bfp_1, \bfp_2)=A^2\,\Pi(\bfp_1,\bfp_2)\,
\Big\{p_1\,p_2\,T(p_1)^2\,T(p_2)^2
+
|\bfp_1+\bfp_2|\,T(|\bfp_1+\bfp_2|)^2\Big[p_1\,T(p_1)^2 +p_2\, T(p_2)^2\Big]
\Big\}\;.
\end{equation}
Again, $\Sigma_{\de}^{(a)}$ corresponds to the $p_1p_2T(p_1)^2T(p_2)^2$
term and $\Sigma_{\de}^{(b)}$ is the remaining contribution. Also
these integrals may be reduced to simpler expressions analytically: the
final formulae can be found in Appendix~D.\\

\begin{figure}
\setlength{\unitlength}{1cm}
\centering
\begin{picture}(6,8)
\put(0.0,0.0){\includegraphics{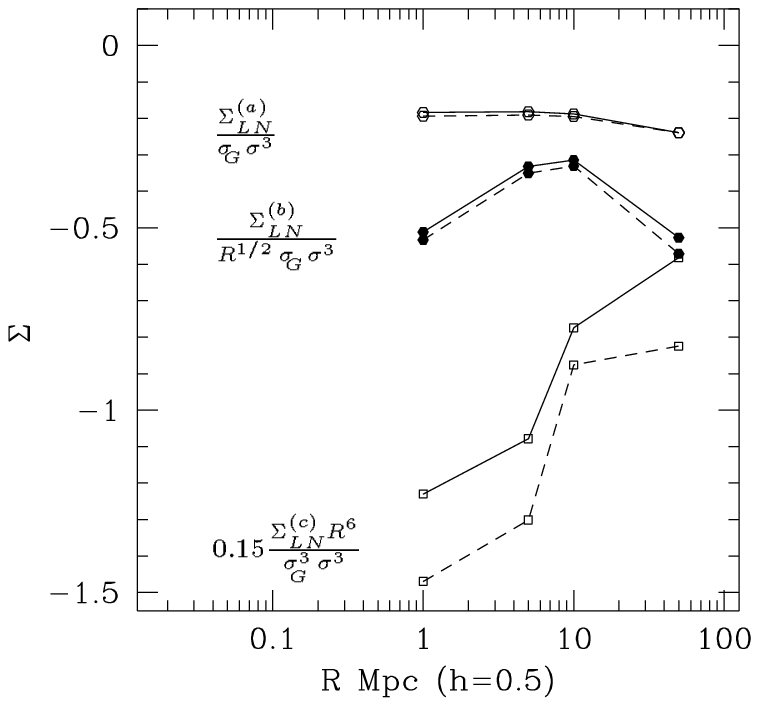}}
\put(0.0,0.0){\includegraphics{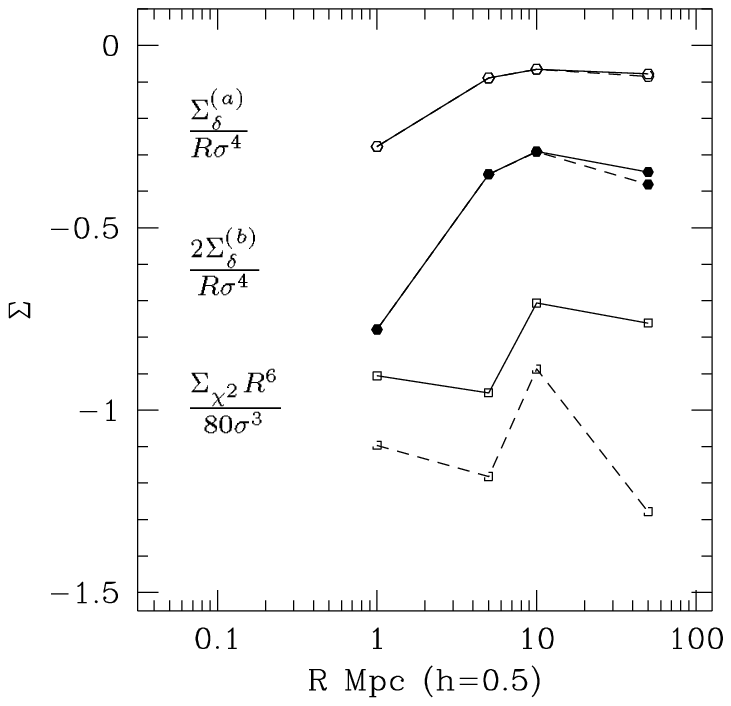}}
\end{picture}
\caption{Correction terms to the linear angular momentum versus scale
$R$, for several non-Gaussian statistics indicated in the panels (only
the positively skewed ones are shown). The $\Sigma$--terms have been
scaled by powers of the mass-variance $\sigma(R)$ to make them
independent of the spectral normalisation and by powers of $R$ to
reduce the scale dependence of the plotted quantity. The lognormal
terms have in addition been normalised by powers of $\sig$. Full lines
corresponds to the choice $p_m=0.005$Mpc$^{-1}$ and dashed lines to
$p_m=0.01$Mpc$^{-1}$; $p_M=2.5$Mpc$^{-1}$ for both, the scale $R$ is
in Mpc ($h=0.5$).}
\label{fig:sigmas}
\end{figure}

Numerical values for the various $\Sigma$s are shown in
Fig.~\ref{fig:sigmas}. Only positively skewed contributions are shown,
negatively skewed have opposite sign. Some terms are logarithmically
divergent for $p_m\rightarrow 0$, i.e., for {\it large} scales. This
divergence can be traced back to the bispectrum. Note that the limits
$p_m< p_{1,2}< p_M$ follow from restricting the initial scale-free
power spectrum $P_G$ to a finite interval; however, this generally sets
no restrictions on $|\bfp_1+\bfp_2|$ and the divergence occurs for
elongated triangles $\bfp_1\rightarrow -\bfp_2$ which can be controlled
by the infra-red cut-off $p_m$ in $\Pi(\bfp_1,\bfp_2)$. There are no
ultra-violet divergences ($p_M\rightarrow \infty$) because such small
scale interactions have been explicitly smoothed over. The different
$\Sigma$s have been scaled by appropriate powers of the mass variance
$\sigma(R)$ to make them independent of the amplitude $A$ of the power
spectrum. In addition, the LN corrections have been scaled by their
dependence on the dispersion $\sig$ of the underlying Gaussian
model. Finally, some $\Sigma$s have been scaled by powers of $R$ to
reduce the scale dependence of the plotted quantity. The
$\Sigma^{(c)}_{LN}$ and $\Sigma_{\chi^2}$ contribution follow from 6D
Monte-Carlo integrations and are less accurate than the others, which
involve 3D numerical integrations.

\begin{figure}
\setlength{\unitlength}{1cm}
\centering
\begin{picture}(6,8)
\put(0.0,0.0){\includegraphics{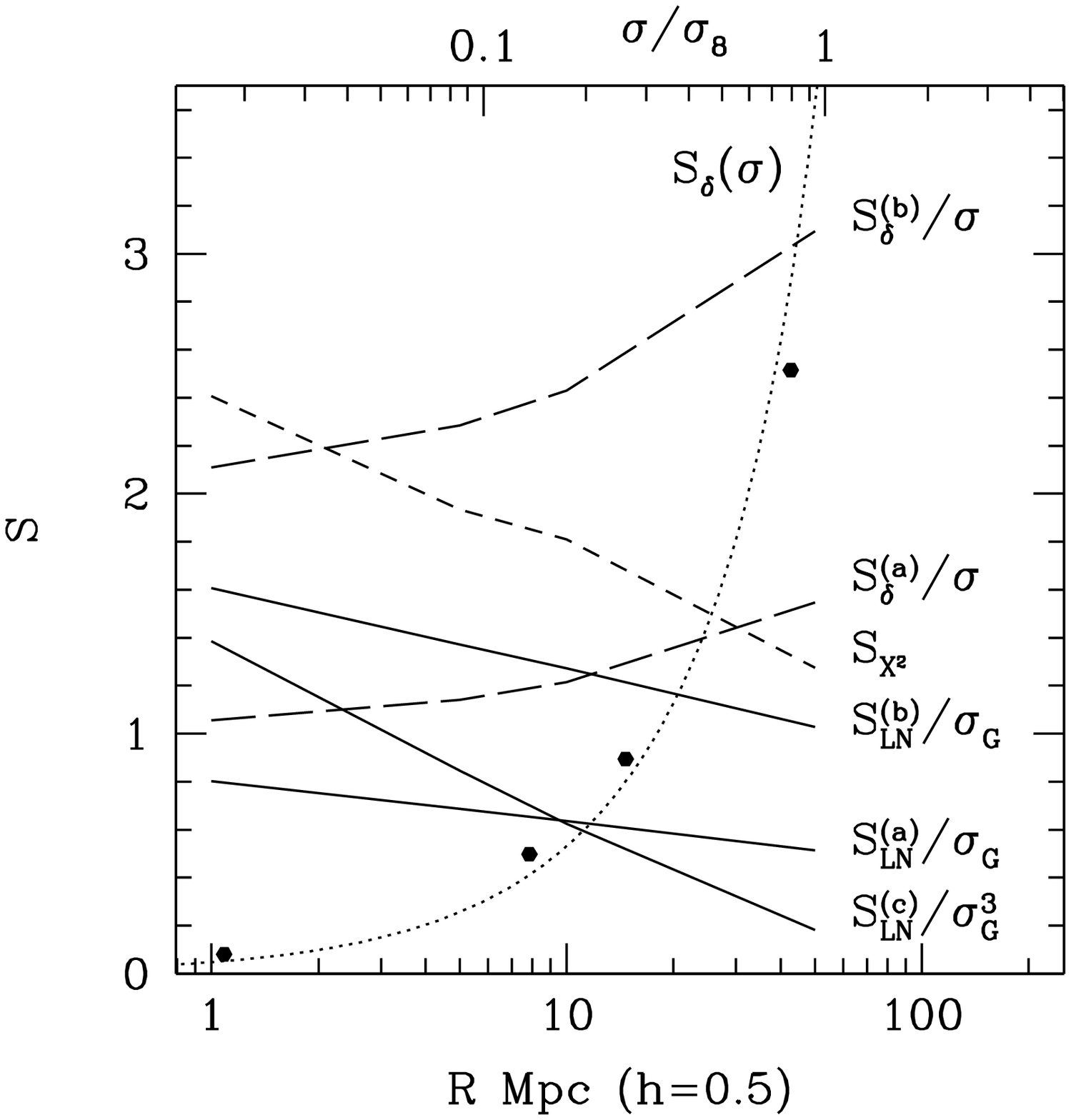}}
\put(0.0,0.0){\includegraphics{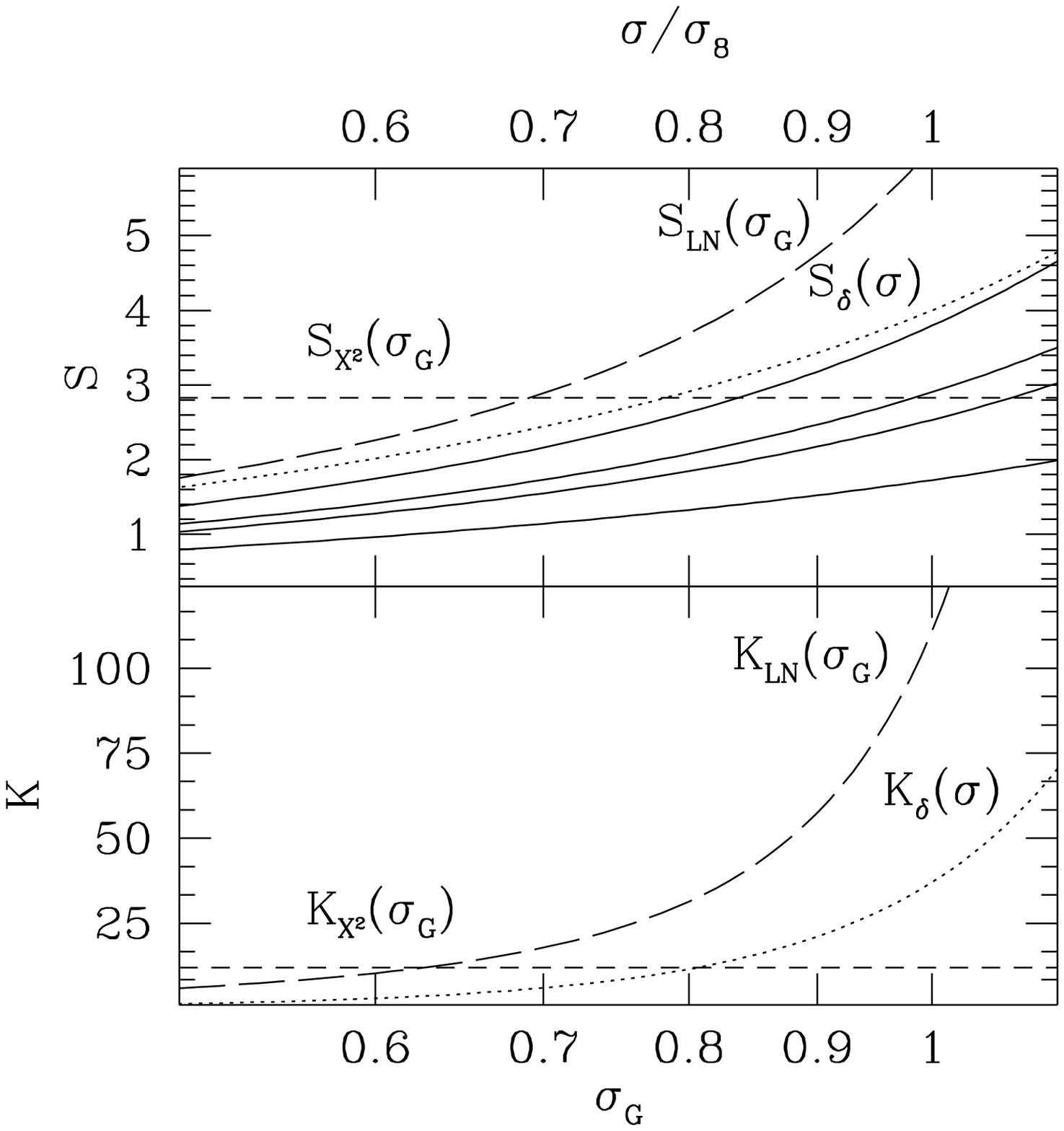}}
\end{picture}
\caption{Left panel, lower scale: skewness $S$ as a function of scale
$R$ (in Mpc, $h=0.5$) for the different (positively skewed) statistics:
$S_{LN}$ and $S_\delta$ denote the density skewness for the lognormal
gravitational potential and the lognormal density field,
respectively. The individual contributions to $S_{LN}$ are scaled by
powers of $\sig$, as indicated; $S_\delta$ is scaled by $\sigma$. The
dotted line (upper scale) is $S_\delta$ versus $\sigma/\sigma_8$ for
the LN density field taken from equation~(43a) in Bernardeau \& Kofman
(1995). Filled diamonds show our determination of
$S^{(a)}_\delta+S^{(b)}_\delta$ versus $\sigma/\sigma_8$ for
comparison. Right panels, lower scale: skewness $S$ (upper panel) and
kurtosis $K$ (lower panel) for the $\chi^2$ (short dashed) and LN (long
dashed) unfiltered models as function of $\sig$. The $\chi^2$ values
are independent of $\sig$. The {\it filtered} skewness for the LN
gravitational potential distribution is plotted for comparison (upper
panel, full lines) for filtering scales (from top to bottom) 1, 5, 10
and 50~Mpc, with $h=0.5$. The dotted lines (upper scale) show the
functions in equations~(43a) and (43b) from Bernardeau \& Kofman (1995)
for skewness (upper panel) and kurtosis (lower panel) versus
$\sigma/\sigma_8$ for the LN density field. For all curves,
$p_m=0.05$Mpc$^{-1}$, $p_M=2.5$Mpc$^{-1}$, $h=0.5$.}
\label{fig:kurtosis}
\end{figure}

\section{Angular momentum at the maximum expansion time}
In this section we quantify the relative non-Gaussian perturbative corrections
to the linear angular momentum for the different statistics by computing the
parameter
\begin{equation}
\Upsilon(R)\equiv {\lan \bfL^2\ran\over\lan \bfL^{(1)^2}\ran} -1 
= 2\, {\lan\bfL^{(1)}\cdot\bfL^{(2)} \ran\over\lan \bfL^{(1)^2}\ran}\;,
\end{equation}
at the maximum expansion time, $D(\tau_R)\sigma(R)=1$ (we computed the
maximum expansion time for the spherical model; see Peebles 1969;
see also Paper~I).\\

\noindent $\bullet$ For the Chi--squared distribution we obtain
\begin{equation}
\Upsilon_{\chi^2}(R) = {2\,\dot E\over \dot D D}\,{\Sigma_{\chi^2}(R)\over
\sigma(R)^3}\,[D(\tau_R)\sigma(R)] \approx -{12\over 7}\,{\Sigma_{\chi^2}(R)\over
\sigma(R)^3}\sim  \pm\,2.1\,(R/h^{-1}{\rm Mpc})^{-6}\;,
\end{equation}
where the last value is estimated from Fig.~(\ref{fig:sigmas}). This
correction is of $O(1)$ on galactic scales. We suggest that this
indicates that higher-order terms in the perturbative expansion
probably contribute significantly. We reiterate that the upper (lower)
sign refers to positively (negatively) skewed distributions.\\

\noindent $\bullet$ For the LN distribution we obtain
\begin{equation}
\Upsilon_{LN}(R) = {2\,\dot E\over \dot D D}\,{\Sigma^{(a)}_{LN}(R)
+\Sigma^{(b)}_{LN}(R)+\Sigma^{(c)}_{LN}(R)\over \sigma(R)^3}\,[D(\tau_R)\sigma(R)]
\sim \pm [0.4\,\sig+\sig (R/h^{-1}{\rm Mpc})^{1/2}
+0.2\sig^3 (R/h^{-1}{\rm Mpc})^{-6}],
\label{63}
\end{equation}
which depends explicitly on $\sig$, i.e., on the level of
non-Gaussianity imposed on the initial state. Below we estimate an
upper limit for $\sigma_G$ and hence an upper limit to this
correction.\\

\noindent $\bullet$ For the linearised LN density field, the scaling is
rather different. This is because the degree of non-Gaussianity is
determined directly by the amplitude of the power-spectrum and
consequently the non-Gaussian correction term must also depend on that
amplitude, parametrised for example by the mass-variance in spheres of
$R_8=8h^{-1}{\rm Mpc}$ at the present time, $\sigma_8$. This
complicates the dependence of the correction on the mass scale
$M$. Approximating the scaling of $\sigma$ with $R$ for the standard
CDM model as $\sigma(R)/\sigma_8\approx (R_8/R)^{1.1+0.25\log(R/R_8)}$
and given the fact that the relevant $\Sigma$s scale $\propto
R\sigma(R)^4$, we obtain
\begin{equation}
\Upsilon_{\delta}(R) ={2\,\dot E\over \dot D
D}\,{\Sigma^{(a)}_{\delta}(R) +\Sigma^{(b)}_{\delta}(R)\over
\sigma(R)^3}\,[D(\tau_R)\sigma(R)]
\approx {2\,\dot E\over D\,\dot
D}\,{\Sigma^{(a)}_{\delta}(R) +\Sigma^{(b)}_{\delta}(R)\over
R\sigma(R)^4}\, R\,\sigma_8\,\left({R_8\over
R}\right)^{1.1+0.25\log(R/R_8)}
\sim \pm 6.3\,\sigma_8\;,
\end{equation}
where the numerical value is appropriate for scales $R=0.5 h^{-1}
$~Mpc. Clearly, LN non-Gaussianity imposed directly on the density
field can change quite drastically the amount of angular momentum
induced on galactic scales. The absolute value of the correction
depends on the normalisation of the power spectrum, through
$\sigma_8$, which contrasts with the other distributions which are
independent of $\sigma_8$.

\subsection{Estimating the degree of non-Gaussianity}
The relative spin correction $\Upsilon_{LN}$ for the case of a
lognormally distributed gravitational potential $\varphi$ depends on
the free parameter $\sigma_G$, cfr equation~({\ref{63}}), which fixes
the degree of non-Gaussianity. An upper limit on the allowed degree of
non-Gaussianity, and hence on $\sigma_G$, can be obtained by analysing
the deviations from Gaussianity for the density field. This also allows
us to show that distributions with similar skewness and kurtosis can
nevertheless induce widely different corrections to the angular
momentum.

In this spirit, we have computed numerically the skewness
$S=\lan\delta^3\ran/\lan\delta^2\ran^{3/2}$ as a function of scale for
all the considered non-Gaussian models starting from their respective
bispectra: the result is shown in Fig.~(\ref{fig:kurtosis}) which also
shows the unfiltered kurtosis
$K=\lan\delta^4\ran/\lan\delta^2\ran^{2}-3$. The dotted lines denote
the expressions given by Bernardeau \& Kofman (1995) for unfiltered
skewness and kurtosis versus $\sigma$ for the LN density field (their
equations~(43a) and (43b)). From Fig.~(\ref{fig:kurtosis}) (left panel)
it is clear that the skewness decreases weakly with scale for all
models, except for the ``linearised'' LN density field which decreases
more strongly: for $\sigma\simlt 1$, $S_{LN}\approx 3\sigma$ (and
$K_{LN}\approx 16\sigma^2$). Note that our estimate for the skewness
for the {\it filtered} ``linearised'' LN density model versus $\sigma$
agrees well with the value for the unfiltered one.

We can estimate an upper limit to $\sig$ for the gravitational
potential LN model by requiring that the primordial skewness of the
density field, $S_{\varphi_{_{LN}}}$, is much smaller than the observed
skewness, $S\approx 4$ on scales $R\approx 8\,h^{-1}$ Mpc. We thus set
\begin{equation}
S_{\varphi_{_{LN}}}(\sig) \approx 4\alpha\;,
\label{skew1}
\end{equation}
with $\alpha\ll 1$. Using the approximations
$S_{\varphi_{_{LN}}}(\sig)\approx 1.8\sig+0.6\sig^3$ (from Fig.~2, for $\sig\ll 1$)
we get
\begin{equation}
\sig\approx 2.22\alpha-3.65\alpha^3+18\alpha^5\;\;\;\;{\rm for\;\;}\alpha << 1\;.
\label{skew2}
\end{equation}
Taking $\alpha\sim 0.1$, we obtain $\sig\sim 0.2$ from which
$\Upsilon_{LN}(R= 0.5h^{-1}{\rm Mpc})\sim \pm 0.33$. We conclude that
in this approximation the non-Gaussian correction to the linear
angular momentum for a LN potential is of the order of $\le 33$ per
cent.\\

Summarising, it is clear that different underlying statistics give rise
to differing corrections to the linear estimate of the angular
momentum. For instance, the contribution of higher-order spin
corrections appears to be non-negligible for $\chi^2$ and for the
``linearised'' LN density field, whereas the perturbative expansion for
the LN potential field seems to converge. Clearly, more detailed
investigations of these non-Gaussian corrections seem warranted.
Finally, note that in all cases positively skewed distributions
increase the angular momentum of collapsing objects with respect to the
linear estimate, and vice versa for negatively skewed ones.

\section{Summary and conclusions}
In this paper we analysed the corrections to the linear growth of the
tidal angular momentum $\bfL$ acquired by a proto-object (protogalaxy
or protocluster) as a consequence of non-Gaussian initial
conditions. We have used Lagrangian perturbation theory for the
displacement field to obtain perturbative corrections to the linear
angular momentum. Whereas the linear rms angular momentum involves
integrals over the power spectrum alone -- and hence is independent of
the assumed statistics of the random field -- the lowest-order
perturbative corrections involve integrals over the bispectrum and so
do depend on the underlying probability distribution. We showed that
for a generic non-Gaussian distribution the corrections to the variance
of the angular momentum grows $\propto t^{8/3}$, which contrasts with
the Gaussian case for which the growth rate is $t^{10/3}$ (both for the
Einstein-de Sitter universe). This is a consequence of the fact that
the lowest order perturbative spin contribution in the non-Gaussian
case arises from the third moment of the underlying density field,
which is identically zero for a Gaussian field. In this formalism, the
resulting expression for the variance of the spin factorises the same
shape parameter as in the Gaussian case, since
$\lan\bfL^{(1)}\cdot\bfL^{(2)}\ran\propto \mu_1^2-3\mu_2$, where
$\mu_1$ and $\mu_2$ are the first and second invariants of the inertia
tensor of the collapsing object. This result does not depend on the
assumed statistics, since it is a consequence of the isotropy of the
Universe (see Appendix~A).

The spin variance was evaluated explicitly as a function of scale for a
variety of multiplicative non-Gaussian statistics, namely $\chi^2$ and
lognormal distributions imposed on the {\it gravitational potential}
and ``linearised'' lognormal statistics assumed directly for the {\it
density} field. We characterised these corrections in terms of
$\Upsilon(R)\equiv \lan \bfL^2\ran/\lan \bfL^{(1)^2}\ran -1$, i.e., in
terms of the relative contribution of non-Gaussian corrections to the
linear spin. In general, during the mildly non-linear regime,
positively skewed distributions increase the angular momentum with
respect to the linear term, and vice versa for negatively skewed ones.
However, the convergence properties of the perturbative series depends
strongly on the assumed statistics. We found that the $\chi^2$ and
``linearised'' lognormal density statistics induce $O(1)$ corrections
to the linear angular momentum, presumably indicating that higher-order
terms, not discussed in this paper, are non-negligible. Consequently,
angular momentum acquisition in models with this kind of underlying
statistic appear analytically intractable. In contrast, the
investigation of the lognormal distributions imposed on the {\it
gravitational potential} seems more promising in this respect, since
the lowest-order non-Gaussian spin correction are of order 30 per cent,
suggesting convergence of the perturbative series. The calculations in
this paper do not take into account non-gravitational processes or
highly non-linear events which may play important roles in the
formation of galaxies.

\section*{Acknowledgements}
We acknowledge Dick Bond, George Efstathiou, Sabino Matarrese, Ravi
Sheth, Radek Stompor and Rien van de Weygaert for discussions. Sabino
Matarrese is particularly thanked for allowing to reproduce in
Appendix~B some of his unpublished notes. PC is personally grateful to
Michal Chodorowski, Ewa {\L}okas and Lauro Moscardini for permitting us
to use some preliminary results of a non--Gaussian collaboration still
in progress, and for their patience for the continuous delays caused to
the common projects. Constructive remarks of the Referee improved this
work. PC and TT were supported by the EEC Human Capital and Mobility
Programme under contracts CT930328 and CT941463 respectively.

{}

\section*{Appendix~A}
In this first appendix we report the details of the derivation of
equation~(\ref{21}) for the spin correction
$\lan\bfL^{(1)}\!\cdot\bfL^{(2)}\ran$, namely
\begin{equation}
\lan\bfL^{(1)}\cdot\bfL^{(2)}\ran = \f{2}{15}\,
\dot{D}(\tau)\dot{E}(\tau)\,(\mu_1^2-3\mu_2)\,\Sigma(R)\;,
\end{equation}
where the wave-vector contribution $\Sigma(R)$ is defined by
\begin{equation}
\Sigma(R) \equiv 15\,\lan\calD^{(1)}_{xy}\,\calD^{(2)}_{xy}\ran =
15\,\calK^{(2)}_{xyxy}\oplus B_\psi^{(1)}\;.
\end{equation}
We will show that this expression is a direct consequence of the
isotropy of the universe.\\

Our starting point is the general expression [see equation~(\ref{18})]
\begin{equation}
\lan\bfL^{(1)}\cdot\bfL^{(2)}\ran=
\dot{D}(\tau)\,\dot{E}(\tau)\,\eps_{\al\beta\gamma}\,\eps_{\al\beta'\gamma'}\,
\calJ_{\s\gamma}\,\calJ_{\s'\gamma'}\,
\lan\calD^{(1)}_{\beta\s}\,\calD^{(2)}_{\beta'\!\s'}\ran\;.
\end{equation}
The Levi-Civita tensors may be written as
$\eps_{\al\beta\gamma}\,\eps_{\al\beta'\gamma'}=
\de_{\beta\beta'}\de_{\gamma\gamma'}-\de_{\beta\gamma'}\de_{\beta'\gamma}$,
which leads to:
\begin{equation}
\lan\bfL^{(1)}\cdot\bfL^{(2)}\ran=
\dot{D}(\tau)\,\dot{E}(\tau)\,\calJ_{\s\gamma}
\big[\calJ_{\s'\gamma}\,
\lan\calD^{(1)}_{\beta\s}\,\calD^{(2)}_{\beta\s'}\ran-
\calJ_{\s'\!\beta}\,
\lan\calD^{(1)}_{\beta\s}\,\calD^{(2)}_{\gamma\s'}\ran\big]\;.
\end{equation}
Without loss of generality, we now assume that the chosen frame of
reference coincides with the eigenframe of the inertia tensor, in which
case $\calJ_{\al\beta}=\iota_\al\de_{\al\beta}$, where $\iota_\al$ are
the eigenvalues of $\calJ$ (no summation over $\alpha$ in this specific
case!). In this particular reference frame, we get
\begin{eqnarray}
\lan\bfL^{(1)}\cdot\bfL^{(2)}\ran&=&
\dot{D}(\tau)\,\dot{E}(\tau)\,(\iota_\s^2 - \iota_\s\iota_\beta)\,
\lan\calD^{(1)}_{\beta\s}\,\calD^{(2)}_{\beta\s}\ran 
\;\;\;\;\;\;\;\;\;\;\;\;\;\;(\beta\neq\s)
\nonumber \\
&=&\dot{D}(\tau)\,\dot{E}(\tau)\,
\big[2(\iota_x^2+\iota_y^2+\iota_z^2)\,
\lan\calD^{(1)}_{xy}\,\calD^{(2)}_{xy}\ran-
2(\iota_x\iota_y+\iota_x\iota_z+\iota_y\iota_z)
\,\lan\calD^{(1)}_{xy}\,\calD^{(2)}_{xy}\ran\big]
\nonumber \\
&=&2\dot{D}(\tau)\,\dot{E}(\tau)\,(\mu_1^2-3\mu_2)
\,\lan\calD^{(1)}_{xy}\,\calD^{(2)}_{xy}\ran\;,
\label{eq:general}
\end{eqnarray}
where $\mu_1\equiv\iota_x+\iota_y+\iota_z$ and 
$\mu_2\equiv\iota_x\iota_y+\iota_x\iota_z+\iota_y\iota_z$ are the first
and the second invariant of the inertia tensor $\calJ$. The latter
equation~(\ref{eq:general}) was used in Section~3.2

\section*{Appendix~B} 
In this Appendix we describe how to obtain the expression of the
Chi--squared power spectrum in equation~(\ref{32}), using the same
notation as in Section~3.1. The power spectrum of the 2--point
correlation function of the field $\vphi = \vphi_\circ\,G^2$ is given
by
\begin{equation}
P_\vphi(p) = 2\vphi_\circ^2 \int
d\bfq\,[\xi_G(q)]^2\, {\rm e}^{-i\bfq\cdot\bfp}
=\f{\vphi_\circ^2}{4\pi^3}\,\int d\bfk\,P_G(k)\,P_G(|\bfk-\bfp|)\;.
\end{equation}
Changing variable from azimuthal angle $\theta$ (using spherical
coordinates in $\bfk$) to $\ell=|\bfk-\bfp|$
simplifies the previous integral to:
\begin{equation}
P_\vphi(p)= \f{\vphi_\circ^2}{2\pi^2\,p}\int_0^\infty dk\,k\,P_G(k)
\,\int_{|p-k|}^{p+k}d\ell \,\ell\, P_G(\ell)\;.
\end{equation}
Let us now consider the flicker--noise power spectrum for $G$ defined
in equation~(\ref{31}) and introduce the parameters $\epsilon \equiv k_m/p$
and $Q\equiv k_M/p$. The condition $k_m\leq p \leq k_M$ then
translates into $\epsilon\leq 1$ and $Q\geq 1$ and the previous
integral simplifies further to 
\begin{equation}
P_\vphi(p) = \f{\vphi_\circ^2}{2\pi^2\,p^3}\int_\epsilon^Q dk\,k^{-2} \,\int_{{\rm
Max}\{\epsilon, |k-1|\}}^{{\rm Min}\{Q, k+1\}}d\ell \,\ell^{-2}\;.
\end{equation}
In the limit $Q \rightarrow \infty$ one has
\begin{equation}
P_\vphi(p) =
\f{\vphi_\circ^2}{2\pi^2\,p^3} \left[ \int_\epsilon^1
\f{dk}{k^2}\f{1}{{\rm Max}\{\epsilon, 1-k\}} + \int_1^\infty
\f{dk}{k^2}\f{1}{{\rm Max}\{\epsilon, k-1\}}-
\int_\epsilon^\infty\f{dk}{k^2}\f{1}{k+1} \right] \;,
\end{equation}
and two expressions for $P_\vphi$ may be recovered: for $p\geq
2\,k_m$, i.e. $\epsilon\leq 1/2$,
\begin{equation}
P_\vphi(p)=\f{\vphi_\circ^2}{2\pi^2\,p^3}\,
2\ln\left(\f{1-\epsilon^2}{\epsilon^2}\right)\;,
\end{equation}
while for $k_m\leq p\leq 2k_m$, i.e. $\epsilon\geq 1/2$, the power
spectrum reduces to
\begin{equation}
P_\vphi(p)= \f{\vphi_\circ^2}{2\pi^2\,p^3}\, \left[
2\ln\left(\f{1+\epsilon}{\epsilon}\right) +
\left(\f{1-\epsilon}{\epsilon}\right)^2 -1 \right]\;.
\end{equation}
Substituting $\epsilon=k_m/p$ into the latter equations leads to the
compact expression for the Chi--squared power spectrum in
equation~(\ref{32}).

\section*{Appendix~C}
In this third Appendix we summarise those properties of the
Chi--squared and lognormal distributions which are relevant to our
analysis. To indicate the non--Gaussian field, we employ the same
symbol $\vphi$ used in the main text for the primordial gravitational
potential; however, the results derived below hold for any
non-Gaussian Chi--squared or lognormal field. The reader interested in
a more comprehensive treatment is addressed to Kendall \& Stuart
(1977).\\

\noindent {\it C.1 ~~Chi--squared distribution}\\
\noindent Let $G$ be a Gaussian variable with zero mean and variance
$\sig^2$. The variable $\vphi \equiv \vphi_\circ\,G^2$ has a
Chi-squared distribution, with characteristic function
\begin{equation}
\Phi_\vphi(t)=\f{1}{\sqrt{1-2i\,\vphi_\circ\,\sig^2\,t\,}}\;.
\end{equation}
The moments $\mu_n$ of $\vphi$ about the origin $\vphi=0$ can be
obtained by evaluating successive derivatives of the characteristic
function $\Phi_\vphi(t)$ at $t=0$:
\begin{equation}
\mu_n = \lan\vphi^n\ran =\left.
(-i)^n\,\f{d^n\Phi_\vphi(t)}{dt^n}\right|_{t=0}\;.
\end{equation}
Explicitly, the first few moments are found to be:
\begin{equation}
\mu_1 =
\vphi_\circ\,\sig^2\;;\;\;\;\;\;\;\; \mu_2 =
3\,\vphi_\circ^2\,\sig^4\;; \;\;\;\;\;\;\;\mu_3
=15\,\vphi_\circ^3\,\sig^6\;.
\label{eq-chimom}
\end{equation}
The reduced moments about the mean, or cumulants, are ($\kappa_1=0$)
\begin{equation}
\kappa_2 = \lan(\vphi-\lan\vphi\ran)^2\ran = \mu_2- \mu_1^2 = 
2\,\vphi_\circ^2\,\sig^4\;, 
\label{70}
\end{equation}
\begin{equation}
\kappa_3 = \lan(\vphi-\lan\vphi\ran)^3\ran= \mu_3 - 3\,\mu_1\,\mu_2 
+ 2\,\mu_1^3 = 8\,\vphi_\circ^3\,\sig^6\;,
\label{71}
\end{equation}
The cumulants $\kappa_n$ with $n>2$ vanish, by definition, for a
Gaussian distribution. The deviation from non-Gaussianity of a field
$\vphi$ with given dispersion $\sigma^2_\vphi$ obtained in this way
cannot be tuned by changing $\sig$ of the underlying Gaussian
field. Indeed, since $\sigma^2_\vphi=3\vphi_0^2\sig^4$ is assumed
to be constant, we find $\kappa_2=2\sigma^2_\vphi/3$ and
$\kappa_3=8\sigma^3_\vphi/3$, independent of $\sig$. Similar
relations hold true for all higher moments as well. The same remark
applies to all fields $\vphi\propto G^n$, obtained from powers of an
underlying Gaussian field. In contrast, we will show below that for LN
statistics, the deviation from Gaussianity can be tuned by varying
$\sig$.

The characteristic function of $\vphi$ is useful to calculate the
moments of its one-point PDF. To compute the {\it correlation
functions} of the field $\vphi$, it is convenient to work with the
characteristic function $\Phi_G^{(n)}$ of the multivariate Gaussian
probability distribution of the field $G(\bfq)$, where the dependence
of $G$ on the `spatial' variable $\bfq$ is now emphasized. It is given
by:
\begin{equation}
\Phi_G^{(n)}(\bft) = \exp \left[-\f{1}{2}\, 
\bft^T \calC \,\bft \right] \;, 
\label{72}
\end{equation}
where $\bft\equiv (t_1,\ldots,t_n)$, $\bft^T$ denotes the transpose of
$\bft$ and $\calC$ is the covariance matrix of the Gaussian field
$G$, whose elements are
\begin{equation}
\lan G(\bfq_i)\, G(\bfq_j)\ran = \xi_G(q_{ij})\;
\label{73}
\end{equation}
and $q_{ij}\equiv|\bfq_i-\bfq_j|\,.$ Here, $\xi_G(q)$ is the
covariance (correlation) function of the Gaussian field $G$ with~
$\xi_G(0) = \sig^2$. By definition, the 2-point correlation function
of the field $\vphi$ is
\begin{equation}
\xi_\vphi(q)\equiv\lan(\vphi_1-\lan\vphi\ran)\,
(\vphi_2-\lan\vphi\ran)\ran = 
\lan\vphi_1 \,\vphi_2\ran  - \lan\vphi\ran^2\;,
\label{74}
\end{equation}
where $\vphi_i\equiv\vphi(\bfq_i)$.
We have
\begin{equation}
\vphi_\circ^{-2}\lan\vphi_1\,\vphi_2\ran = \lan G_1^2\, G_2^2\ran = 
\left.\f{\p^4\Phi_G^{(2)}}{\p t_1^2\,\p t_2^2}
\right|_{\bft={\bf 0}} = 
\sig^4 + 2\,\xi_G(q)^2\;. 
\label{75}
\end{equation}
Combining equations~(\ref{74}) and (\ref{75}), we obtain the following
relation between the correlation function of $\vphi$ and that of $G$:
\begin{equation}
\xi_\vphi(q) = 2\,\vphi_\circ^2\,\xi_G(q)^2\;.
\label{76}
\end{equation}
In a similar way, using the characteristic function of order $n$ of
the multivariate Gaussian distribution function, one may derive the
higher-order correlation functions of the field $\vphi$.  Thus, the
3-point connected correlation function is
\begin{equation}
\zeta_\vphi(123)\equiv\lan(\vphi_1-\lan\vphi\ran)\,
(\vphi_2-\lan\vphi\ran)\, 
(\vphi_3-\lan\vphi\ran)\ran = 
8\,\vphi_\circ^3\,\xi_G(12)\,\xi_G(13)\,\xi_G(23)\;,
\label{77}
\end{equation}
where $\xi_G(ij)\equiv\xi_G(\bfq_i, \bfq_j)$, and so on. Note that,
from equations~(\ref{76}) and (\ref{77}), one has $\kappa_2 =
\xi_\vphi(0) = 2\vphi_\circ^2\,\sig^4$ and $\kappa_3 = \zeta_\vphi(0)=
8\vphi_\circ^3\,\sig^6$, in agreement with equations~(\ref{70}) and
(\ref{71}). The bispectrum $B_\vphi$ may be recovered by performing
the Fourier transform of equation~(\ref{77}), which gives the
expression reported in equation~(\ref{33}). We stress once more the
absence of any explicit dependence on the underlying variance
$\sig^2$.\\

%
\noindent{\it C.2 ~~Lognormal distribution} \\
%
\noindent Let us again suppose that $G$ is a normal variable with zero
mean and variance $\sig^2$. The variable $\vphi\equiv
\vphi_\circ\,{\rm e}^G$ is said to be lognormal distributed. The
2-point correlation function of the field $\vphi$, defined in equation
(\ref{74}), may be obtained from the correlation $\xi_G$ of the field
$G$ via the bivariate characteristic $\Phi_G^{(2)}$.  From
equations~(\ref{72}) and (\ref{73}) we get
\begin{equation}
\vphi_\circ^{-2}\,\lan \vphi_1\,\vphi_2\ran=
\lan{\rm e}^{\,G_1+G_2}\ran= 
\Phi_G^{(2)}(-i, -i) = {\rm e}^{\,\sig^2}\,{\rm e}^{\xi_G(q)} \;.
\end{equation}
Moreover, $ \vphi_\circ^{-1}\,\lan\vphi\ran = 
\Phi_G^{(1)}(-i) = {\rm e}^{\,\sig^2/2}$. 
Combining these expressions one gets again a relation between the two
2-point correlation functions:
\begin{equation}
\xi_\vphi(q) =\lan\vphi\ran^2\, \left[{\rm e}^{\,\xi_G(q)}-1\right] \;.
\label{eq-log2corr}
\end{equation}
In a similar fashion, one can obtain the 3--point correlation
function, defined in equation (\ref{77}). 
Since
\begin{equation}
\zeta_\vphi(123)=
\lan\vphi_1\,\vphi_2\,\vphi_3\ran-\lan\vphi\ran\,\lan\vphi_1\,\vphi_2\ran-
\lan\vphi\ran\,\lan\vphi_2\,\vphi_3\ran-
\lan\vphi\ran\,\lan\vphi_1\,\vphi_3\ran+2\,\lan\vphi\ran^3 \;,
\label{eq-3cor}
\end{equation}
one obtains
\begin{equation}
\lan\vphi_1\vphi_2\vphi_3\ran=
\lan\vphi\ran^3 + 
\lan\vphi\ran\left[
\xi_\vphi(12)+\xi_\vphi(13)+\xi_\vphi(23)\right]+\zeta_\vphi(123)\;.
\label{eq-3corcom}
\end{equation}
On the other hand, the left hand side of the latter equation may be written 
as
\begin{eqnarray}
\lan\vphi_1\,\vphi_2\,\vphi_3\ran&=&
\vphi_\circ^3\,\lan{\rm e}^{\,G_1+G_2+G_3}\ran= 
\vphi_\circ^3\,\Phi_G^{(3)}(-i, -i, -i) \nonumber \\
&=&\lan\vphi\ran^3 \,{\rm e}^{\,\xi_G(12)+\xi_G(13)+\xi_G(23)} =
\lan\vphi\ran^3 \,
[1+\lan\vphi\ran^{-2}\xi_\vphi(12)]\,
[1+\lan\vphi\ran^{-2}\xi_\vphi(13)]\,[1+\lan\vphi\ran^{-2}\xi_\vphi(23)]
\nonumber \\
&=&
\lan\vphi\ran^{-3}\,\xi_\vphi(12)\,\xi_\vphi(13)\,\xi_\vphi(23)+
\lan\vphi\ran^{-1}\left[
\xi_\vphi(12)\,\xi_\vphi(13)+\xi_\vphi(12)\,\xi_\vphi(23)+ 
\xi_\vphi(13)\,\xi_\vphi(23)\right]\nonumber \\
&+&\lan\vphi\ran\,\left[\xi_\vphi(12)+\xi_\vphi(13)+\xi_\vphi(23)\right]
+\lan\vphi\ran^3\;,
\label{eq-3char}
\end{eqnarray}
and, from (\ref{eq-3corcom}), one recovers a Kirkwood-like expansion
for the 3-point connected correlation function:
\begin{equation}
\zeta_\vphi(123)= \lan\vphi\ran^{-1}\left[
\xi_\vphi(13)\,\xi_\vphi(23)+\xi_\vphi(12)\,\xi_\vphi(13)+ 
\xi_\vphi(12)\,\xi_\vphi(23)\right]+ 
\lan\vphi\ran^{-3}\left[\xi_\vphi(12)\,\xi_\vphi(13)\,\xi_\vphi(23)\right]\;.
\label{83}
\end{equation}
The first few cumulants $\kappa_n$ of this lognormal distribution are:
\begin{equation}
\kappa_2 = \xi_\vphi(0) = \vphi_\circ^2\,{\rm e}^{\,\sig^2}\,
\left({\rm e}^{\,\sig^2}-1\right)\;,
\end{equation}
\begin{equation}
\kappa_3 = \zeta_\vphi(0) = \vphi_\circ^3\, {\rm e}^{3\sig^2/2}\,
\left({\rm e}^{\,\sig^2}-1\right)^2
\left({\rm e}^{\,\sig^2}+2\right) \;.
\end{equation}
Taking again $\sigma_\vphi^2=\kappa_2$ constant by setting
$\vphi_0^2=\kappa_2/e^{\sig^2}\,(e^{\sig^2}-1)$ where $\sig$ is now a
free parameter, one can write
$\kappa_3=\kappa_2^{3/2}\,(e^{\sig^2}+1)(e^{\sig^2}-1)^{1/2}$, from
which it is clear that the skewness of $\vphi$ can be tuned
independently of its variance. Clearly, this is also true for higher
moments. In other words, the deviation from Gaussianity of the LN
$\vphi$ can be tuned by changing $\sig$. In particular, in the limit
$\sig^2\rightarrow 0$ the lognormal distribution tends to a
Gaussian. The bispectrum $B_\vphi$ of the lognormal variable $\vphi$
may be derived by Fourier transforming the expression (\ref{83})
(Catelan, Chodorowski, {\L}okas and Moscardini 1997). The final result
is reported in equation~(\ref{38}). An explicit dependence on the
underlying variance $\sig^2$ can be noted in equation~(\ref{40}),
which holds in the regime $\sig^2 \simlt 1$.

\section*{Appendix~D}
In this last appendix we want to show how to perform analytically some
of the integrations when computing the spin correction $\Sigma(R)$ for
the different non-Gaussian fields we have considered. The compact
expression for $\Sigma(R)$ in terms of the kernel $\calK^{(2)}_{xyxy}$
enables us to reduce easily the number of numerical integrations by
exploiting rotational invariance of the integrals.\\

\noindent $\bullet$ We start by considering the simplest case, 
the $\Sigma_{LN}(R)$ correction for the LN gravitational potential $\vphi$:
\begin{equation}
\Sigma_{LN}(R) = 15\,\calK^{(2)}_{xyxy}\oplus B_\psi^{(1)}=
\Sigma_{LN}^{(a)}(R)+\Sigma_{LN}^{(b)}(R)+\Sigma_{LN}^{(c)}(R)\;.
\end{equation}
As discussed in the main text, the three contributions
$\Sigma_{LN}^{(h)}(R)$ originate from the Kirkwood relation of the LN
bispectrum $B_\psi^{(1)}$ [see equations (\ref{29}) and (\ref{38})].
Let us consider the first of these three contributions:
\begin{equation}
\Sigma_{LN}^{(a)}(R)=
15\,\lan\vphi\ran^{-1}\,\calK^{(2)}_{xyxy}(\bfp_1,\bfp_2)
\oplus \calT(p_1)\,\calT(p_2)
\,\calT(|\bfp_1+\bfp_2|)\,P_\vphi(\bfp_1) \,P_\vphi(\bfp_2)\;.
\end{equation}
which can be written explicitly as:
\begin{equation}
\Sigma_{LN}^{(a)}(R)=
-\f{15}{\lan\vphi\ran}\!\int\!\!\f{d\bfp_1\,d\bfp_2}{(2\pi)^6}\,
\f{(\bfp_1+\bfp_2)_x^2\,(\bfp_1+\bfp_2)_y^2\,}
{|\bfp_1+\bfp_2|^2}\,\kappa^{(2)}(\bfp_1, \bfp_2)\,
[\fW(|\bfp_1+\bfp_2|R)]^2 
\,\calT(p_1)\,\calT(p_2)\,
\calT(|\bfp_1+\bfp_2|)\,P_\vphi(\bfp_1) \,P_\vphi(\bfp_2)\;.
\end{equation}
We now use the relation $P_\vphi(p)\approx A_\vphi
F_\vphi(p)p^{-3}\Pi(p)$, with $F_\vphi=1$ for the LN case (and
hence $\calT(p)=T(p)$) and in addition $A_\vphi=2\pi^2
\sig^2\,\lan\vphi\ran^2/\ln(p_M/p_m)\approx A$. Furthermore, without
loss of generality, we assume that the vector $\bfp_1$ is chosen along
the $z$-direction, in which case the previous expression simplifies
to:
\begin{equation}
\Sigma_{LN}^{(a)}(R)=-15\,\f{A_\vphi^2}{\lan\vphi\ran}\,
\int \f{d\bfp_1d\bfp_2}{(2\pi)^6}\,
\f{p_{2x}^2\,p_{2y}^2}{|\bfp_1+\bfp_2|^2}\, \kappa^{(2)}(\bfp_1,
\bfp_2)\, \big[\fW(|\bfp_1+\bfp_2|R)\big]^2\,T(p_1)\,
T(p_2)\,T(|\bfp_1+\bfp_2|)\,p_1^{-3}\,p_2^{-3}\Pi(\bfp_1,\bfp_2)\;.
\end{equation}
Note that the integrand does not depend on the azimuthal angle
$\theta_1$ nor on the precessional angle $\phi_1$; in addition, the
$\phi_2$-integration may be done analytically. The final result may be
cast in the form:
\begin{eqnarray}
\Sigma_{LN}^{(a)}(R)=&-&\!\!\!\f{15}{8(2\pi)^4}\,\left[\pm A^{3/2}
\sqrt{\f{2\pi^2\sig^2}{\ln(p_M/p_m)}\,}\,\right] \int_0^\infty
dp_1\,T(p_1)
\int_0^\infty dp_2\, p_2^4\,T(p_2) \int_0^{\pi}d\theta_2\,({\rm sin}\theta_2)^7\,
\nonumber \\ &\times& \!\!\!\big(p_1\,p_2^{-1} + p_2\,p_1^{-1}+2{\rm
cos}\theta_2\big)^{-1}\,
T(|\bfp_1+\bfp_2|)\,\big[\fW(|\bfp_1+\bfp_2|R)\big]^2\Pi(\bfp_1,\bfp_2)\;,
\end{eqnarray}
where $\theta_2$ is the angle between $\bfp_1$ and $\bfp_2$. The upper
(lower) sign is for positively (negatively) skewed LN distributions. The
remaning integral needs to be computed numerically.\\

The other two contributions may be calculated in a similar way.
The second contribution is
\begin{eqnarray}
\Sigma_{LN}^{(b)}(R)=&-&\!\!\!\f{15}{4(2\pi)^4}\,\left[\pm A^{3/2}
\sqrt{\f{2\pi^2\,\s^2_G}{\ln(p_M/p_m)}\,}\,\right]
\int_0^\infty dp_1\,p_1^{-3/2}\,T(p_1)\int_0^\infty dp_2\, 
p_2^{11/2}\,T(p_2)
\int_0^\pi d\theta_2\,({\rm sin}\theta_2)^7\,
\nonumber \\
&\times&
\!\!\!\big(p_1\,p_2^{-1} + p_2\,p_1^{-1}+2{\rm cos}\theta_2\big)^{-5/2}\,
T(|\bfp_1+\bfp_2|)\,\big[\fW(|\bfp_1+\bfp_2|R)\big]^2\Pi(\bfp_1,\bfp_2)\;.
\end{eqnarray}
The computation of the third contribution is the more involved, 
since it corresponds to a 9 dimensional integration. Recalling that
the bispectrum $B_\vphi$ depends only on the moduli of the vectors
$\bfp_1$ and $\bfp_2$ and the mutual angle $\theta_{12}=\theta_2$, one
has
\begin{eqnarray}
\Sigma_{LN}^{(c)}(R)=&-&\!\!\!\f{15}{8(2\pi)^7}\,
\left[\pm A^{3/2}
\left(\f{2\pi^2\,\s^2_G}{\ln(p_M/p_m)}\right)^{3/2}\,\right]
\int_0^\infty dp_1\,p_1^{3/2}\,T(p_1)\int_0^\infty  dp_2\, 
p_2^{11/2}\,T(p_2)
\int_0^\pi d\theta_2\,({\rm sin}\theta_2)^7\,
\nonumber \\
&\times&
\!\!\!\big(p_1\,p_2^{-1} + p_2\,p_1^{-1}+2{\rm cos}\theta_2\big)^{-1}\,
T(|\bfp_1+\bfp_2|)\,\big[\fW(|\bfp_1+\bfp_2|R)\big]^2\,
\calB(p_1, p_2; \theta_2)\;,
\end{eqnarray}
where
\begin{eqnarray}
\calB(p_1, p_2; \theta_2)&\equiv&
\int_0^\infty dp_3\,p_3^{-4}\int_0^\pi d\theta_3\,{\rm sin}\theta_3\,
\big(p_1\,p_3^{-1} + p_3\,p_1^{-1}+2{\rm cos}\theta_3\big)^{-3/2}
\nonumber \\
&\times&\int_0^{2\pi} d\phi_3
\left[p_2\,p_3^{-1}+p_3\,p_2^{-1}-2(
{\rm cos}\theta_3{\rm cos}\theta_2+{\rm sin}\theta_3{\rm sin}\theta_2
{\rm cos}\phi_3)\right]^{-3/2}\,\Pi(\bfp_1,\bfp)\Pi(\bfp_2,-\bfp)
\Pi(\bfp_1,\bfp_2)\;.
\end{eqnarray}
This leaves a 6D numerical integration to be performed.\\

\noindent$\bullet$ The Chi-squared correction $\Sigma_{\chi^2}(R)$ may
be calculated in the same fashion as the contribution
$\Sigma_{LN}^{(c)}(R)$; the result is
\begin{eqnarray}
\Sigma_{\chi^2}(R)=&-&\!\!\!\f{15}{(2\pi)^4}\,
\left[\pm \left(\f{A}{2}\right)^{3/2}\right]
\int_0^\infty dp_1\,p_1^{3/2}\,\calT(p_1)\int_0^\infty dp_2\, 
p_2^{11/2}\,\calT(p_2)
\int_0^\pi d\theta_2\,({\rm sin}\theta_2)^7\,
\nonumber \\
&\times&
\!\!\!\big(p_1\,p_2^{-1} + p_2\,p_1^{-1}+2{\rm cos}\theta_2\big)^{-1}\,
\calT(|\bfp_1+\bfp_2|)\,\big[\fW(|\bfp_1+\bfp_2|R)\big]^2\,
\calB(p_1, p_2; \theta_2)\,\Pi(\bfp_1,\bfp)\Pi(\bfp_2,-\bfp)\;,
\end{eqnarray}
where $\calT(p)\equiv T(p)[\beta(p)+2\ln(1+p/p_m)-1]^{-1/2}$.\\

\noindent $\bullet$ For the ``linearised'' LN density field one can 
perform calculations in a similar way, which lead to equation (\ref{56}),
where the function 
$\Sigma_{\de}(R)=\Sigma_{\de}^{(a)}(R)+\Sigma_{\de}^{(b)}(R)$ is the sum 
of two contributions originated by the two terms of the linear bispectrum 
$B_\de^{(1)}$ in equation~(\ref{50}):
\begin{eqnarray}
\Sigma_{\de}^{(a)}(R)=&-&\!\!\!\f{15}{8}\,\f{A^2}{(2\pi)^4}
\int_0^\infty dp_1\,p_1\,T(p_1)^2\int_0^\infty dp_2\, 
p_2^5\,T(p_2)^2
\int_0^\pi d\theta_2\,({\rm sin}\theta_2)^7\,\nonumber\\
&\times& \big(p_1\,p_2^{-1} + p_2\,p_1^{-1}+2{\rm cos}\theta_2\big)^{-2}\,
\big[\fW(|\bfp_1+\bfp_2|R)\big]^2\Pi(\bfp_1,\bfp_2)\;.
\end{eqnarray}
The second contribution is
\begin{eqnarray}
\Sigma_{\de}^{(b)}(R)=&-&\!\!\!\f{15}{8}\,\f{A^2}{(2\pi)^4}
\int_0^\infty dp_1\,p_1^{3/2}\,T(p_1)^2\int_0^\infty dp_2\, 
p_2^{9/2}
\int_0^\pi d\theta_2\,({\rm sin}\theta_2)^7\,
\nonumber \\
&\times&
\!\!\!\big(p_1\,p_2^{-1} + p_2\,p_1^{-1}+2{\rm cos}\theta_2\big)^{-3/2}\,
\big[T(|\bfp_1+\bfp_2|)\big]^2\,\big[\fW(|\bfp_1+\bfp_2|R)\big]^2
\Pi(\bfp_1,\bfp_2)\;.
\end{eqnarray}

\end{document}